\documentclass[reprint,twocolumn,superscriptaddress, amssymb,aps,prl]{revtex4-1}
\usepackage{graphicx}
\usepackage{xcolor} 
\usepackage{amsmath}
\usepackage{hyperref}

\newcommand{\ket}{\right \rangle}

\begin{document}
\title{Raman Sideband Cooling of Molecules in an Optical Tweezer Array}
\author{Yukai Lu}
\thanks{These authors contributed equally to this work.}
\affiliation{Department of Physics, Princeton University, Princeton, New Jersey 08544 USA}
\affiliation{Department of Electrical and Computer Engineering, Princeton University, Princeton, New Jersey 08544 USA}
\author{Samuel J. Li}
\thanks{These authors contributed equally to this work.}
\affiliation{Department of Physics, Princeton University, Princeton, New Jersey 08544 USA}
\author{Connor M. Holland}
\thanks{These authors contributed equally to this work.}
\affiliation{Department of Physics, Princeton University, Princeton, New Jersey 08544 USA}
\author{Lawrence W. Cheuk}
\email{lcheuk@princeton.edu}
\affiliation{Department of Physics, Princeton University, Princeton, New Jersey 08544 USA}

\date{\today}

\begin{abstract}

Ultracold molecules, because of their rich internal structures and interactions, have been proposed as a promising platform for quantum science and precision measurement. Direct laser-cooling promises to be a rapid and efficient way to bring molecules to ultracold temperatures. For trapped molecules, laser-cooling to the quantum motional ground state remains an outstanding challenge. A technique capable of reaching the motional ground state is Raman sideband cooling, first demonstrated in trapped ions and atoms. In this work, we demonstrate for the first time Raman sideband cooling of molecules. Specifically, we demonstrate 3D Raman cooling for single CaF molecules trapped in an optical tweezer array, achieving average radial (axial) motional occupation as low as $\bar{n}_r=0.27(7)$ ($\bar{n}_z=7.0(10)$). Notably, we measure a 1D ground state fraction as high as 0.79(4), and a motional entropy per particle of $s = 4.9(3)$, the lowest reported for laser-cooled molecules to date. These lower temperatures could enable longer coherence times and higher fidelity molecular qubit gates desirable for quantum information processing and quantum simulation. With further improvements, Raman cooling could also be a new route towards molecular quantum degeneracy applicable to many laser-coolable molecular species including polyatomic ones. 

\end{abstract}

\maketitle

Ultracold molecules have been proposed as a new platform for exploring many areas in physics ranging from simulation of quantum many-body Hamiltonians, to quantum information processing, to precision measurements in searches for physics beyond the Standard Model~\cite{demille2002quantum, Carr2009review,Bohn2017molreview,blackmore2018reviewMoleculeQuantum}. Yet, cooling and fully controlling molecules have been long-standing experimental challenges. One route to produce ultracold molecules is via assembly from atoms, for which cooling techniques are well-developed. This approach has successfully been used to produce bialkali molecules, enabling explorations in ultracold chemistry~\cite{DeMiranda2011StereoChem,Liu2021UCChem} and the creation of quantum-degenerate molecular gases well-suited for studying long-ranged interacting many-body systems~\cite{demarco2019degKrb,schindewolf2022NaKEvap}. 

In contrast to assembly from ultracold atoms, methods that directly cool could be broadly applicable to a large variety of molecular species including polyatomic ones. In particular, direct laser-cooling of molecules has seen great advances recently. Starting with molecular magneto-optical traps near the Doppler limit of $\sim 100\,\mu\text{K}$~\cite{barry2014SrFMOT,Truppe2017Mot,Anderegg2017MOT,Collopy2018MOT,Vilas2022MOTCaOH}, sub-Doppler cooling techniques have allowed experiments to enter into the $\mu\text{K}$ regime~\cite{Truppe2017Mot,Cheuk2018Lambda,Caldwell2019CaFSD,Ding2020YOSD,Vilas2022MOTCaOH,Langin2021SrFODT,Wu2021YOODT,hallas2022CaOHODT}. Importantly, sub-Doppler cooling has enabled optically trapped molecular samples with record densities~\cite{Anderegg2018ODT,Cheuk2018Lambda,Wu2021YOODT} and arrays of single molecules trapped in optical tweezers~\cite{Anderegg2019Tweezer,Holland2022Tweezer,Bao2022Tweezer}.

Access to molecular samples at even lower temperatures could enable new possibilities. For example, molecular tweezer arrays have recently emerged as a promising platform for quantum science. Notably, recent work has shown high-fidelity control over molecular positions and internal states, coherent dipolar interactions, and implementation of an entangling two-qubit gate~\cite{Holland2022Tweezer,Bao2022Tweezer}. Yet, residual thermal motion limits the achievable coherence times and gate fidelities. These limitations can be largely eliminated by cooling to the motional ground state. 

One technique capable of cooling to the motional ground state is Raman sideband cooling (RSC)~\cite{Heinzen1990RSCproposal}. First, a Raman process transfers a molecule initially in internal state $\left|\uparrow\ket$ to $\left|\downarrow\ket$ while removing $\Delta n$ quanta of motional energy. Subsequently, optical pumping reinitializes the internal state to $\left|\uparrow\ket$ while largely preserving the motional state. By iterating over these two steps, cooling is achieved. RSC was first proposed and demonstrated for trapped ions and atoms in optical lattices~\cite{Monroe1995RSC,Hamann1998RSC,Kerman2000RSC}, and has since been used to cool single atoms in tweezer traps to their motional ground states~\cite{Kaufman2012RbRSC,Thompson2012RbRSC}, for imaging in quantum gas microscopes~\cite{Cheuk2015qgm,Parsons2015site}, and to produce single molecules via assembly from two RSC-cooled atoms~\cite{He2020RbRbTweezer,Zhang2021NaCsArray,Ruttley2023RbCsTweezer}. RSC also provides a rapid and efficient (low-loss) all-optical method to create atomic Bose-Einstein condensates~\cite{Hu2017RSCBEC,Urvoy2019DRSCtoBEC}, circumventing the need for evaporation.

\begin{figure}[h!]
	{\includegraphics[width=\columnwidth]{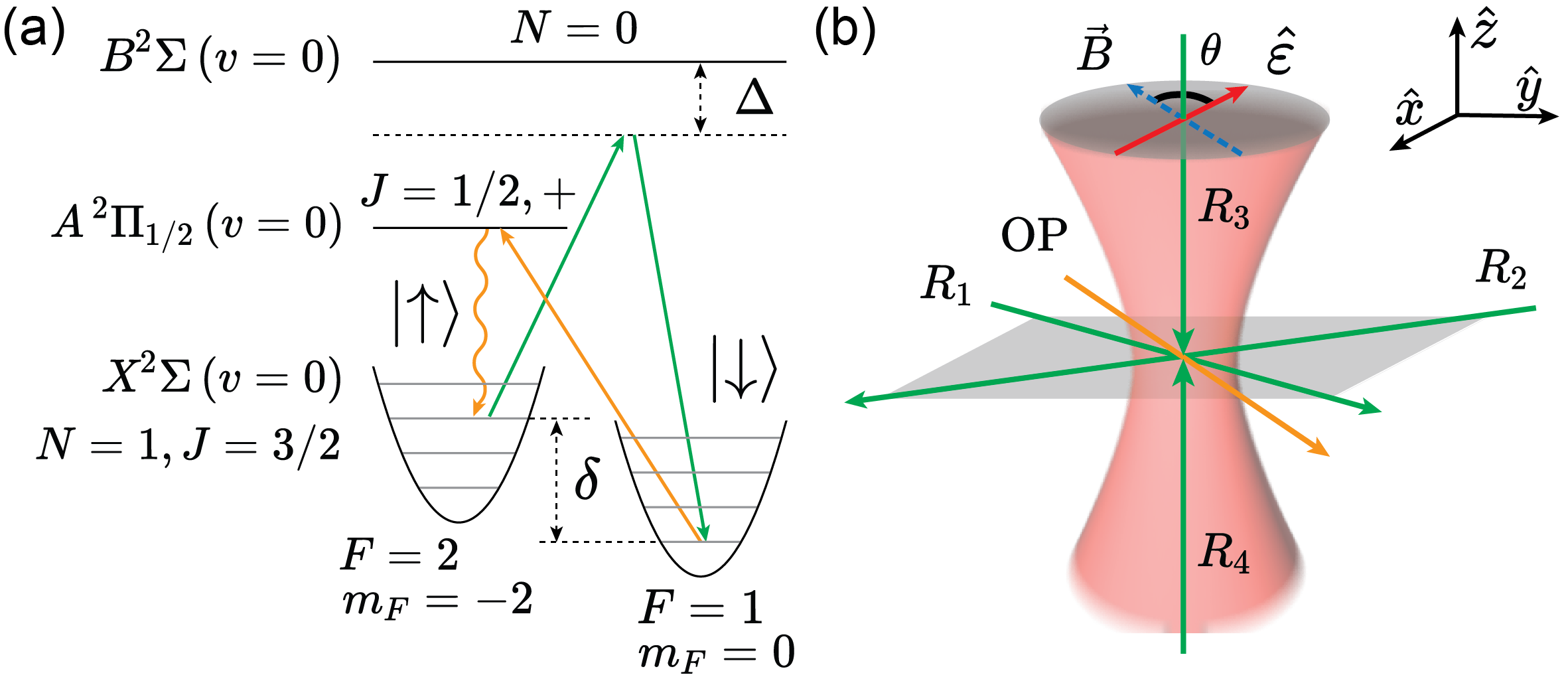}}
	\caption{\label{Fig_1} Raman Sideband Cooling Scheme. (a) Motional-changing two-photon Raman transitions between $\left|\uparrow\ket$ and $\left|\downarrow\ket$ are driven using laser beams detuned by $\Delta\approx-2\pi \times 42~\text{GHz}$ from the $X^2\Sigma(v=0, N=1) \to B^2\Sigma(v=0, N=0)$ transition. $\delta$ denotes the two-photon detuning. Optical pumping into $\left|\uparrow\ket$ is performed using light addressing the $X^2\Sigma(v=0, N=1) \to A^2\Pi_{1/2}(v=0, J=1/2,+)$ transition. (b) A magnetic field $\vec{B}$ is applied in the radial plane, and is oriented at an angle of $\theta$ relative to the polarization axis of the tweezer light $(\hat{\varepsilon}\parallel \hat{x})$. Raman laser beams $R_1, R_2$ ($R_3,R_4$) address the radial (axial) directions. $R_1$ and $R_2$ are optionally retro-reflected. Optical pumping light (OP) is applied radially.}
	\vspace{-0.2in}
\end{figure}
Raman sideband cooling of optically trapped and laser-cooled molecules faces two main challenges that arise from the complex internal structure of molecules~\cite{caldwell2020sideband}. First, state-dependent optical trapping leads to inhomogeneous broadening of Raman transitions, preventing resolved addressing of cooling sidebands and decreasing transfer efficiency. Second, efficient optical pumping is difficult because of the large number of molecular states and the degradation of free-space selection rules in deep optical traps.

In this work, we demonstrate Raman sideband cooling of molecules for the first time. We overcome the above challenges by devising a RSC scheme for CaF molecules that provides both narrow Raman transitions and efficient optical pumping. Our scheme does not require high magnetic fields, in contrast to the one proposed in~\cite{caldwell2020sideband}. We demonstrate 3D Raman cooling for CaF molecules trapped in an optical tweezer array and achieve average motional quanta as low as $\bar{n}_r=0.27(7)$ and $\bar{n}_z=7.0(10)$ in the radial and axial directions, respectively.

Our work begins with single CaF molecules that are cooled by $\Lambda$-enhanced gray molasses~\cite{Cheuk2018Lambda} and trapped in a 1D array of linearly polarized optical tweezers~\cite{Anderegg2019Tweezer,holland2022bichromatic,Holland2022Tweezer}. Raman beams are sent along the radial and axial directions, and are near-detuned from the $X^2\Sigma(v = 0, N = 1)\rightarrow B^2\Sigma(v =0, N = 0) $ transition (Fig. \ref{Fig_1}). Optical pumping is achieved on the $X^2\Sigma(v = 0, N = 1)\rightarrow A^2\Pi_{1/2}(v =0, J = 1/2,+)$ transition.

We identify a suitable pair of internal states $\{\left|\uparrow\ket, \left|\downarrow\ket\}$ for RSC. Constrained by optical pumping, we consider optically cyclable states from $X^2\Sigma(v=0, N=1)$. In free space, selection rules enable optical pumping into the stretched states $\left|\pm\ket =\left| N=1, J=3/2, F=2, m_F=\pm2\ket$. Specifically, these states are dark to $\sigma_\pm$ and $\pi$ light addressing the $X^2\Sigma(v=0, N=1) \rightarrow A^2\Pi_{1/2}(v=0,J=1/2,+)$ transition. In deep tweezer traps, the trapping light can admix in bright states and modify selection rules, degrading optical pumping. The admixture can be reduced by providing a well-defined quantization axis with a magnetic field $\vec{B}$ along the polarization axis of the trapping light~\cite{Thompson2012RbRSC,caldwell2020sideband}. However, at certain fields, because of tensor ac Stark shifts, level crossings can occur, increasing bright state admixtures. Our calculations indicate that $\left|\uparrow\ket =\left|-\ket$ is immune from these crossings at low fields and over a large range of trap depths, therefore robustly providing low bright state admixtures~\cite{Supplement}. In particular, the bright state population admixture remains below $10^{-4}$ even for traps as deep as $k_B \times 2\,\text{mK}$ at $B=4.4\,\text{G}$.

For $\left|\downarrow \ket$, we seek a state that is connected by a two-photon Raman transition to $\left|\uparrow\ket$ and has minimal differential ac Stark shifts with $\left|\uparrow\ket$. Our calculations indicate that $\left|\downarrow \ket =\left|N=1, J=3/2,F=1, m_F=0\ket$ satisfies these requirements. In particular, while most $N=1$ states experience fractional differential Stark shifts at the $10^{-1}$ level, those between $\left|\uparrow \ket$ and $\left|\downarrow \ket$ are suppressed to $10^{-2}$ even in deep traps with depths $\sim k_B \times 1\,\text{mK}$. The shifts can be further reduced by changing the angle $\theta$ between $\vec{B}$ and the tweezer polarization, with $10^{-3}$ fractional shifts possible at specific ``magic'' angles. For example, at a trap depth of $k_B \times 0.3\,\text{mK}$ and magnetic field of $B= 5.5~\text{G}$, the magic angle is $\theta \approx 57^\circ$.

\begin{figure}[t!]
	{\includegraphics[width=\columnwidth]{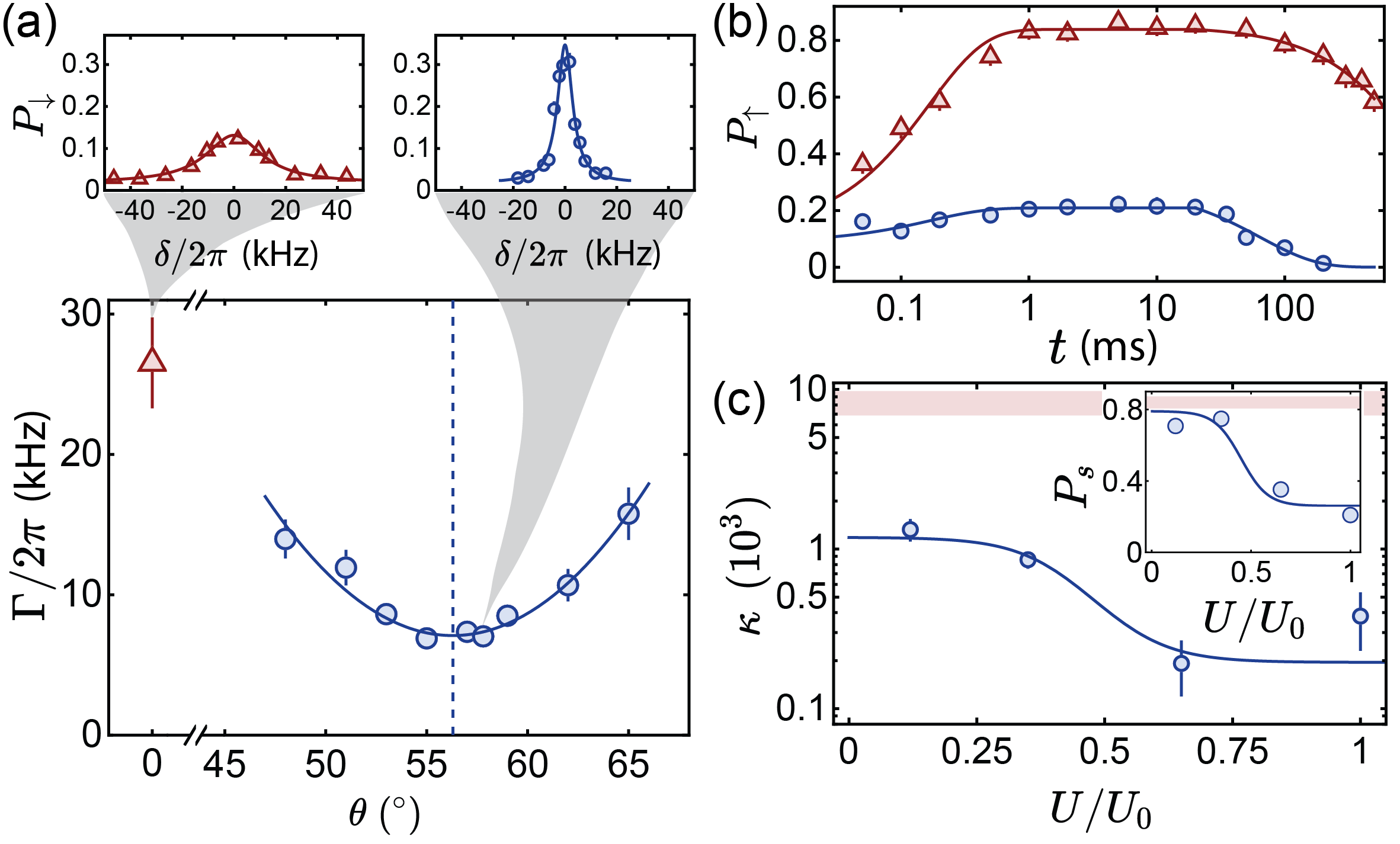}}
	\caption{\label{Fig_2} Raman Linewidths and Optical Pumping Characterization. (a) Raman linewidths $\Gamma$ versus $\theta$, measured at a tweezer depth of $k_B \times  326(7)\,\mu\text{K}$. $\Gamma$ is smallest at the magic angle $\theta_m = 56.3(3)^\circ$, indicated by the dashed vertical line. Representative spectra versus two-photon detuning $\delta$, along with Lorentzian fits (solid), are shown in the sub-panels. (b) $\left |\uparrow\ket$ population $P_\uparrow$ versus optical pumping time $t$ at a tweezer depth of $k_B \times 930(20)\,\mu\text{K}$. Red triangles (blue circles) show data for $\theta=0^\circ$ ($\theta=\theta_m$). Solid curves show simultaneous fits to early and late time dynamics. (c) The figures-of-merit $\kappa$ and $P_s$ at $\theta_m$ versus trap depth $U/U_0$ ($U_0=k_B \times 930(20)\,\mu\text{K}$) are shown by blue circles, with solid lines as guides to the eye. The corresponding values for $\theta=0^\circ$ are shown by the red shaded regions and indicate significantly better optical pumping.}
	\vspace{-0.2in}
\end{figure}

We experimentally verify the properties of $\{\left|\uparrow\ket, \left|\downarrow\ket\}$ through Raman spectroscopy and optical pumping dynamics. We first probe the inhomogeneous broadening of Raman transitions arising largely from differential Stark shifts. At a trap depth $U=k_B \times 326(7)\,\mu\text{K}$, we measure the linewidth $\Gamma$ of the carrier ($\Delta n=0$) transition using co-propagating Raman beams. We prepare molecules in $\left|\uparrow\ket$ and measure the population in $\left|\downarrow\ket$ ($P_\downarrow$) versus the two-photon detuning $\delta$. At $\theta=0^\circ$ and $B=4.4~\text{G}$, we measure a linewidth of $\Gamma=2\pi \times 26.5(3)\,\text{kHz}$. This is below the radial trapping frequency $\omega_r=2\pi \times 117.3(4)\,\text{kHz}$, allowing radial sidebands ($\Delta n\neq0$) to be resolved. At $B=5.5~\text{G}$, as a function of $\theta$, we find that $\Gamma$ reaches a minimum of $\approx 2\pi \times 7\,\text{kHz}$ at the magic angle of $\theta_m = 56.3(3)^\circ$, as predicted (Fig. \ref{Fig_2}(a)). Notably, at $\theta_m$, $\hbar\Gamma/U \approx 10^{-3}$, and the linewidth $\Gamma$ is below the axial trapping frequency $\omega_z \approx 2\pi \times 26\,\text{kHz}$, allowing axial sidebands to be resolved~\cite{Supplement}.
 
Next, we measure optical pumping dynamics. Starting with molecules initially distributed over all 12 hyperfine states of $X^2\Sigma(v=0,N=1)$, we measure the $\left|\uparrow\ket$ population ($P_\uparrow$) after variable durations of optical pumping. At short times, $P_\uparrow$ increases as molecules are pumped into $\left|\uparrow\ket$, and subsequently saturates to $P_s$. At long times, $P_\uparrow$ decreases due to molecular loss arising from heating or decay into undetected states due to imperfect darkness of $\left|\uparrow\ket$ (Fig.~\ref{Fig_2}(b)). The darkness of $\left|\uparrow\ket$, essential for efficient optical pumping, can therefore be quantified by $\kappa = \tau_1/\tau_2$, where $\tau_1$ ($\tau_2$) is the $1/e$ rise (fall) time of $P_\uparrow$. $P_s$ provides a complementary measure of optical pumping efficiency. When $\theta=0^\circ$ and $B=4.4\,\text{G}$, $\kappa \approx 8\times10^3$ and $P_s\approx 0.8$, even at a deep trap depth of $k_B \times 930(20)\,\mu\text{K}$ (Fig.~\ref{Fig_2}(c)). In comparison, in the magic configuration ($\theta=\theta_m$, $B = 5.5~\text{G}$), we find that both $\kappa$ and $P_s$ decrease with increasing trap depth, indicating degrading selection rules. For all tweezer depths explored, $\kappa$ is a factor of 6 to 20 lower compared to that at $\theta=0^\circ$. This shows that optimal optical pumping (at $\theta=0^\circ$) and Raman linewidths (at $\theta=\theta_m$) cannot be simultaneously achieved.

For Raman sideband cooling, deep tweezer depths help preserve the motional state during optical pumping, which is critical for cooling. Although the magic configuration at $\theta_m$ provides the narrowest Raman transitions, optical pumping is severely degraded at deep depths. On the other hand, at $\theta=0^\circ$, optical pumping is efficient even in deep tweezers. The Raman linewidth is slightly broader but sufficient to resolve the radial sidebands. We therefore choose to perform Raman cooling at $\theta=0^\circ$.

\begin{figure}[h!]
	{\includegraphics[width=\columnwidth]{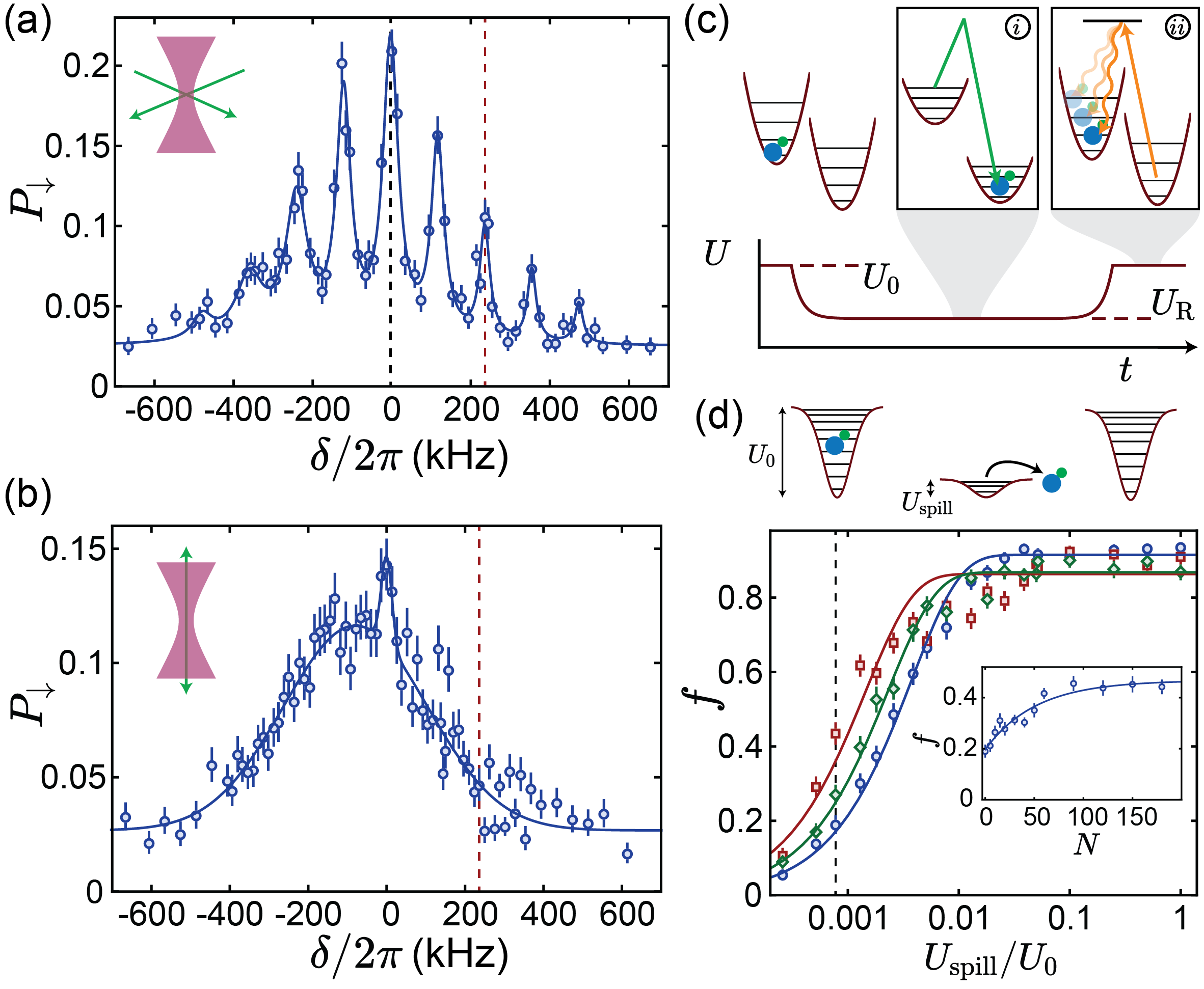}}
	\caption{\label{Fig_3} Raman Sideband Spectra at $\theta=0^\circ$ and Adiabatic Trap Lowering Curves. (a) Population transfer $P_\downarrow$ using radial Raman beams versus $\delta$, the two-photon detuning from the carrier ($\Delta n=0$). The solid blue line shows a fit using a sum of nine Lorentzians with an offset. The black (red) dashed line marks the carrier ($\Delta n_r=-2$ sideband). (b) Population transfer $P_\downarrow$ using axial Raman beams versus $\delta$. Solid curve is a guide to the eye. The red dashed line shows the two-photon detuning $\delta=2\omega_r$ used during cooling. (c) Raman cooling sequence consists of (i) Raman transfers at a tweezer depth of $U_R=k_B \times 326(7)\,\mu\text{K}$, and (ii) optical pumping at a higher depth of $U_0=k_B \times 930(20)\,\mu\text{K}$. (d) Probing temperature via adiabatic trap lowering. The tweezer depth is lowered to $U_{\text{spill}}$ over $1\,\text{ms}$, held $10\,\text{ms}$ to allow hot molecules to escape, and increased back to full depth for detection. The surviving fraction $f$ versus $U_{\text{spill}}/U_0$ is shown for no cooling (blue circles), radial cooling (RC) (green diamonds), and radial/axial cooling (RAC) (red squares). Inset: $f$ versus number of cooling cycles $N$ at a fixed lower depth $U_{\text{spill}}=k_B\times0.72(2)~\mu\text{K}$ (dashed line in main plot).}
	\vspace{-0.2in}
\end{figure}

We next verify Raman motional coupling by driving sideband transitions at $\theta=0^\circ$. The Raman coupling between different motional states is characterized by the Lamb-Dicke parameter $\eta=|\Delta \vec{k}|l/\sqrt{2}$, where $\Delta \vec{k}$ is the difference in wave-vectors between the two Raman beams and $l=\sqrt{\hbar/(m\omega)}$ is the harmonic oscillator length, $m$ being the molecular mass and $\omega$ the trapping frequency. With molecules initially prepared in $\left|\uparrow\ket$, we probe radial and axial motional state transfer at $U=k_B \times 326(7)\,\mu\text{K}$, $B=4.4\,\text{G}$. We pulse on Raman beams and measure $P_\downarrow$ versus $\delta$. The radial Lamb-Dicke parameter is $\eta_r=0.46$ at this depth, allowing us to observe resolved radial sidebands at $\delta = (\Delta n_r )\omega_r$ up to $|\Delta n_r|=4$ (Fig. \ref{Fig_3}(a)), where $\omega_r = 2\pi \times 117.3(4) \,\text{kHz}$. Axially, the weaker confinement leads to a larger Lamb-Dicke parameter of $\eta_z=1.34$, allowing significant motional coupling up to $|\Delta n_z|\sim10$ (Fig. \ref{Fig_3}(b)). We observe that the axial spectrum is significantly broader than the measured carrier Raman linewidth, indicating that motion-changing Raman transfers are indeed occurring. 

Having demonstrated motional state-changing Raman transfer, we next construct a radial cooling sequence consisting of two discrete steps: optical pumping and Raman transfer on a cooling $(\Delta n<0)$ sideband. We optically pump at a deep tweezer depth of $U_0=k_B \times 930(20)\,\mu\text{K}$ to minimize increase in motional quanta. We estimate that $\sim19$ photons are required for optical pumping, increasing the energy by an equivalent of $\Delta n_{r}^{\text{op}}\sim1.3$ radial quanta. To attain net cooling, we therefore address the $\Delta n_r=-2$ sideband. To obtain sufficient motional coupling and reduce inhomogeneous broadening, we perform Raman transfer at a reduced tweezer depth $U_R=k_B \times 326(7)\,\mu\text{K}$, the same depth where we measured Raman linewidths. Each cooling cycle has a duration of $0.65\,\text{ms}$ and the trap depths are ramped adiabatically over $0.2\,\text{ms}$ between $U_0$ and $U_R$ (Fig.~\ref{Fig_3}(c)).

To verify cooling, we first indirectly probe the temperature via adiabatic reduction of the tweezer depth \cite{Cooper2018AEA,tuchendler2008energy}. Hot molecules are spilled progressively as the trap is lowered. At a fixed final trap depth $U_{\text{spill}}$, the surviving fraction $f$ increases with lower temperatures. After 90 cycles of radial cooling (RC), we indeed observe that $f$ increases, indicating radial cooling (Fig. \ref{Fig_3}(d)). We next add axial cooling by switching on additional axial Raman beams with the same two-photon detuning. This radial/axial cooling sequence (RAC) simultaneously addresses  the $\Delta n_r=-2$ radial and the $\Delta n_z \approx -9$ axial sidebands. With 90 cycles of RAC, $f$ increases further compared to radial cooling alone (Fig.~\ref{Fig_3}(d)), indicating successful cooling in all directions. 

The cooling rate of RAC can be probed by measuring $f$ versus the number of cooling cycles $N$. Fixing $U_{\text{spill}}=k_B\times0.72(2)~\mu\text{K}$, we find a $1/e$ cooling timescale of $N_c=51(14)$ cycles. We also quantify loss during cooling. With Raman and optical pumping beams off but keeping the tweezer depth ramps, we observe a $1/e$ lifetime of $1.28(15)\times10^3$ cycles. Remarkably, with Raman cooling on, the lifetime increases to $2.7(4)\times10^3$ cycles, corresponding to a fractional loss of $3.66(6)\times10^{-4}$ per cycle. This difference could arise from Raman cooling compensating for technical heating, or from Raman beams repumping molecules that decay into $X^2\Sigma (v=0,N=3)$ due to off-resonant scattering of tweezer light~\cite{holland2022bichromatic}.

\begin{figure}
	{\includegraphics[width=\columnwidth]{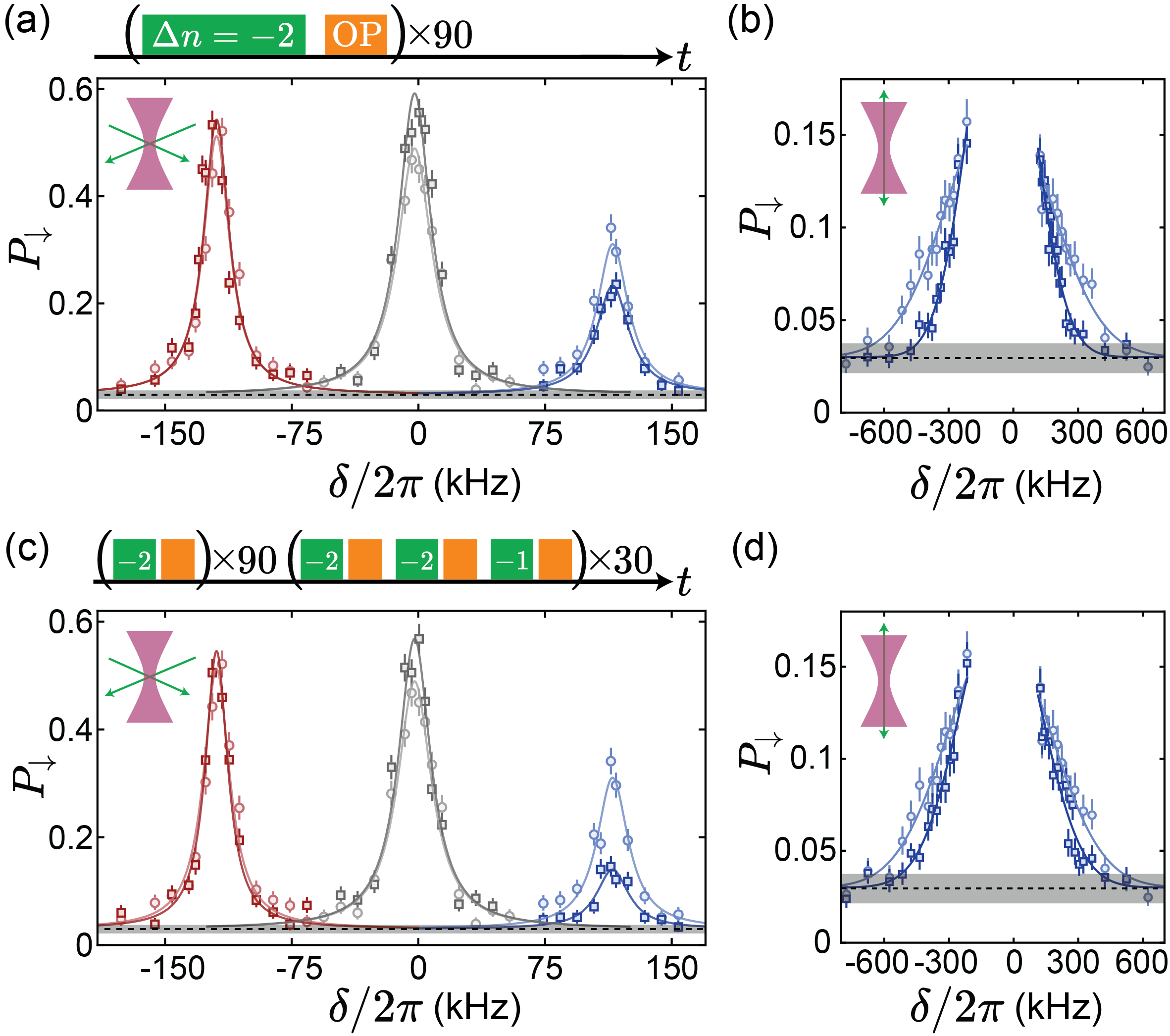}}
	\caption{\label{Fig_4} Raman Thermometry. (a,c) Radial Raman spectra showing the carrier along with $\Delta n_r = \pm1$ sidebands. Grey squares and light gray circles show the carrier after and before cooling, respectively. Red (blue) squares show the $\Delta n_r=1$ ($\Delta n_r=-1$) sideband after cooling; light red (light blue) circles show the $\Delta n_r=1$ ($\Delta n_r=-1$) sideband before cooling. (b,d) Unresolved axial Raman spectra. Blue squares (light blue circles) show spectra after (before) cooling. (a,b) Spectra after 90 cycles of RAC (darker colors) compared to spectra before Raman cooling (lighter colors). (c,d) Spectra after 90 cycles of RAC and 30 cycles of IRC (darker colors) and spectra before Raman cooling (lighter colors). For radial (axial) data, solid lines show fits to Lorentzian (Gaussian) distributions with a vertical offset. For all panels, the dashed line and the shaded region ($\pm1$ standard deviation) show the independently measured offset without Raman beams.}
	\vspace{-0.2in}
\end{figure}

Although adiabatic trap reduction provides qualitative evidence of 3D cooling, it does not directly provide a quantitative temperature. For thermometry, we rely on Raman spectroscopy. We first cool at $\theta=0^\circ$, and then perform spectroscopy in the magic configuration ($\theta=\theta_m$ and $B=5.5\,\text{G}$) to minimize Raman broadening. Along the radial direction, the $\Delta n_r = \pm1$ sidebands are well-resolved. Given the radial Lamb-Dicke parameter and the temperature regime, the Raman motional coupling depends weakly on the motional state $n_r$~\cite{Supplement}. This allows us to observe coherent $\Delta n_r = \pm1$ sideband transfer and also simplifies interpretation of spectra~\cite{Supplement}. We apply $\pi$-pulses and measure the resulting transfer around each sideband (Fig. \ref{Fig_4}(a)). The ratio $\mathcal{A}$ between the peak transfer fractions of the heating versus cooling sidebands allows us to extract 1D ground state fractions and temperatures. Assuming either uniform motional coupling or thermal occupation, we find a 1D radial ground state fraction of $P_0 = 1-1/\mathcal{A} = 0.60(6)$ after radial/axial cooling (RAC). Assuming only a thermal distribution, the mean radial motional occupation $\bar{n}_r$ is given by $\bar{n}_r = 1/(\mathcal{A}-1)$. We find $\bar{n}_r=0.66\,(16)$ ($\tilde{T}_r = k_B T_r/(\hbar \omega_r) = 1.1(2)$) compared to $\bar{n}_r=1.4\,(4)$ ($\tilde{T}_r = 1.8(5)$) before cooling.

We probe the axial temperature similarly. Although we can observe resolved axial sidebands spaced up to $|\Delta n_z|\sim 10$, extracting a temperature is difficult due to the complex lineshape arising from high axial temperatures and a large Lamb-Dicke parameter~\cite{Supplement}. Nevertheless, robust thermometry is possible by probing high-order sidebands in the unresolved regime~\cite{Thompson2012RbRSC,Zhang2022SrClockRSC}. The wings of the spectra become Gaussian, and the spectra can be understood as Doppler-sensitive two-photon transfer~\cite{Supplement}. By fitting the wings, one can robustly extract a temperature. Experimentally, we increase the Raman Rabi coupling such that the wings of the spectra appear smooth (Fig.~\ref{Fig_4}(b)). Fitting the spectra gives an axial temperature of $\tilde{T}_z=k_B T_z/(\hbar\omega_z)=7.5\substack{+1.0 \\ -0.7}$ ($\bar{n}_z=7.0\substack{+1.0\\-0.7}$) after RAC, compared to $\tilde{T}_z=26.5\substack{+4.0 \\ -3.1}$ ($\bar{n}_z=26\substack{+4 \\ -3}$) before cooling.

Lastly, we demonstrate a way to reach even lower temperatures. In our RAC scheme, since $\Delta n_r=-2$ sidebands are addressed, molecules can accumulate in $n_r=1$, limiting the radial ground state fraction. To circumvent this while maintaining net cooling, we apply an additional radial cooling sequence, where we interlace radial cooling cycles that separately address the $\Delta n_r=-2$ and $\Delta n_r=-1$ sidebands (Fig. \ref{Fig_4}(c)). With an additional 30 cycles of interlaced radial cooling (IRC), we observe that the 1D radial ground state fraction increases to $P_0= 0.79(4)$,
corresponding to an average radial occupation of $\bar{n}_r=0.27(7)$ ($\tilde{T}_r=0.65(9)$). Because the axial direction is not cooled during IRC, the axial temperature increases to $\tilde{T}_z=13.5(14)$ ($\bar{n}_z=13.0(14)$), still below the initial temperature. 

When the motion is highly quantized, as in our case, a useful figure-of-merit in addition to temperature is the motional entropy. This quantifies how many motional states are populated and indicates the level of control over the initial motional state. With RAC, we obtain a motional entropy per particle of $s=5.2\substack{+0.5 \\ -0.4}$ compared to $s=7.5\substack{+0.8 \\ -0.7}$ before cooling. With IRC, we further reduce the motional entropy per particle to $s=4.9(3)$, the lowest reported to date for laser-cooled molecules. Since most of the entropy is in the axial motion, lower entropies and temperatures could be reached with increased axial confinement and further optimized Raman pulse sequences~\cite{Yu2018NaCsRSC, Zhang2022SrClockRSC, Spence2022RSC}. 

The motional entropy also allows us to quantify the efficiency of our Raman cooling scheme. In evaporative cooling of atomic and molecular gases, a common efficiency metric is $\gamma= - d\ln(\text{PSD})/d \ln(N)$ where $N$ is the particle number, and $\text{PSD}$ is the phase space density~\cite{Ketterle1996evaporative,Li2021tuning,Bigagli2023collisionally,Lin2023microwave}. One can generalize this metric to $\gamma_q = d s/d\ln(N)$, where $s$ is the motional entropy per particle~\cite{Supplement}. $\gamma_q$ coincides with $\gamma$ in the classical regime and is convenient when in the highly quantized regime. Using our loss measurements, we estimate that $\gamma_q = 70(28)$ for RAC, indicating highly efficient cooling with little loss.

Our demonstration of Raman sideband cooling of molecules in this work opens up several new possibilities. In the near term, the lower temperatures achieved could significantly improve coherence times and provide more coherent dipolar interactions between molecules, enabling high-fidelity quantum gates and quantum simulation with long evolution times. Longer term, our work opens the door to direct laser-cooling of trapped molecules to their 3D motional ground states. This would be a key step towards full quantum control of molecules, and could enable efficient production of ensembles with low motional entropy suited for quantum simulation of itinerant many-body systems. Potentially, this could provide an all-optical route towards quantum degeneracy~\cite{Hu2017RSCBEC,Urvoy2019DRSCtoBEC} that may be broadly applicable to other laser-coolable molecules including polyatomic ones.

\section{Acknowledgements}
We thank Jeff Thompson, Waseem Bakr, and the Bakr group for fruitful discussions. This work is supported by the National Science Foundation under Grant No. 2207518. L.W.C. acknowledges support from the Sloan Foundation. S.J.L. acknowledges support from the Princeton Quantum Initiative. 

\newpage

\section{Methods}
\subsection{Preparation of Molecules}
CaF molecules in the $X^2\Sigma(v=0, N=1)$ manifold are created in a single-stage cryogenic buffer gas cell~\cite{Hutzler2012CBGB}, slowed via chirped slowing, and loaded into a DC magneto-optical trap (MOT). The MOT is subsequently switched off, $\Lambda$-cooling is applied, and the molecules are loaded into an optical dipole trap (ODT) with the aid of a repulsive ring trap in the presence of $\Lambda$-cooling~\cite{Lu2021Ring}. The molecules are optically transported and loaded into a reconfigurable array of 37 optical tweezers, which are created with focused laser beams of 781\,nm light projected vertically through a microscope objective~\cite{holland2022bichromatic}. For normalization, after tweezer loading, the occupation of the tweezers are detected non-destructively with $\Lambda$-imaging~\cite{Holland2022Tweezer}. After non-destructive detection, the molecules are spread out over all 12 hyperfine states in the $X^2\Sigma(v=0, N=1)$ manifold.

\subsection{State-Resolved Detection}
To probe the population in $\left|\downarrow\ket$, we first lower the trap depth to $U_\text{MW}=k_B\times130(3)\,\mu\text{K}$ and rotate the magnetic field ($B= 4.4\,\text{G}$) to $\theta = 53^\circ$.
Microwaves at $\sim20.5$ GHz are used to transfer the population from $\left|\downarrow\ket = \left|N=1, J=3/2, F=1, m_F=0\ket$ to $\left| N=0, J=1/2, F=1, m_F=-1\ket$ via a Landau-Zener sweep. A pulse of light resonant with the $X^2\Sigma(v=0, N=1) \rightarrow A^2\Pi_{1/2}(v=0, J=1/2,+)$ transition removes any molecules remaining in $X^2\Sigma(v=0, N=1)$~\cite{holland2022bichromatic,Holland2022Tweezer}. A second Landau-Zener microwave sweep transfers molecules into $\left|N=1, J=1/2,F=0, m_F=0\ket$. Finally, the population is measured via $\Lambda$-imaging~\cite{Cheuk2018Lambda}, which detects all molecules in $X^2\Sigma(v=0, N=1)$. An analogous approach is used to measure population in $\left|\uparrow\ket$.

\subsection{Optical Pumping}
For optical pumping at $\theta = 0^\circ$, we use a single beam (OP) in the radial (horizontal) plane. The beam makes an angle of $42^\circ$ relative to the tweezer polarization axis $\hat{\varepsilon}$ ($\hat{x}$). It is optimized to have minimal $\sigma_+$ component along the quantization axis set by the magnetic field (along $\hat{x}$ at $\theta=0^\circ$). For the data exploring optical pumping at the magic angle $\theta=\theta_m$ (Fig.~\ref{Fig_2}(b,c)), we use a second optical pumping beam (OP2) propagating along $\hat{y}$. This beam is optimized to have minimal $\sigma_+$ component along the quantization axis set by the magnetic field.

\subsection{Raman Coupling}
To achieve Raman coupling between $\left|\uparrow\ket$ and $\left|\downarrow\ket$, we use light with two frequency components $\omega_{1}$ and $\omega_{2}$ detuned near the $X^2\Sigma(v=0, N=1) - B^2\Sigma(v=0, N=0)$ transition (single-photon detuning of $\Delta=-2\pi\times 42\,\text{GHz}$). $\omega_1$($\omega_2$) nominally couples to $\left|\uparrow\ket$($\left|\downarrow\ket$). These components are generated by acousto-optical modulators (AOMs), allowing the two-photon detuning $\delta$ to be varied.

To achieve motional coupling in the radial (horizontal) $xy$-plane, we send beams $R_1$ and $R_2$ along the $\hat{x} + \hat{y}$ and $\hat{x} - \hat{y}$ directions, respectively. $R_1$($R_2$) carries a single frequency component $\omega_1$($\omega_2$) and is linearly polarized vertically along $\hat{z}$. The two beams are optionally retro-reflected. The retro-reflections are controlled by shutters that can be controlled mid-sequence. To address axial motion, we send beams $R_3$ and $R_4$ along $-\hat{z}$ and $\hat{z}$, respectively. The beam $R_3$ is linearly polarized along $\hat{y}$, while $R_4$ is linearly polarized perpendicular to $\theta=\theta_m$.

\subsection{Spectroscopy Sequences}
We use the following sequence for the data presented in Fig.~\ref{Fig_2}(a).
First, molecules are pumped into $\left|\uparrow\ket$ by turning on OP for $2\,\text{ms}$ in a bias magnetic field of $B=4.4\,\text{G}$ at $\theta = 0^\circ$. Next, the tweezer depth is decreased to $U_R$
and the magnetic field is rotated to the desired angle $\theta$. Both $\omega_{1}$ and $\omega_{2}$ are delivered via a single vertical beam $R_4$ along the tweezer axis. A Raman pulse is applied for $170\,\mu\text{s}$ with an estimated two-photon Rabi coupling of $\Omega_R \approx 2\pi \times 3$ kHz. Finally, the population in $\left|\downarrow\ket$ is measured.

For all spectra in Figs.~\ref{Fig_3} and \ref{Fig_4}, the radial Raman beams $R_1$ and $R_2$ are not retro-reflected. All spectra in Fig.~\ref{Fig_3} are taken in a bias field of $B=4.4\,\text{G}$ at $\theta = 0^\circ$. For the radial spectrum in Fig.~\ref{Fig_3}(a), radial beams $R_1$ and $R_2$ are applied for 1\,ms, probing the radial direction $\hat{y}$. For the axial spectrum in Fig.~\ref{Fig_3}(b), $\omega_{1}$ is delivered via $R_3$ and $\omega_{2}$ via $R_4$, and the beams are applied for 1\,ms.

All spectra in Fig.~\ref{Fig_4} are taken in a magnetic field of $B=5.5\,\text{G}$ at $\theta = \theta_m$. For radial spectra in Fig.~\ref{Fig_4}(a,c), $R_1$ and $R_2$ are used. The beams are applied for $30\,\mu\text{s}$ (carrier) or $50\,\mu\text{s}$ ($\Delta n_r = \pm 1$ sidebands), with $\Omega_R \approx 2\pi \times 20$\,kHz. For axial spectra in Fig.~\ref{Fig_4}(b,d), $\omega_{1}$ is delivered via $R_3$ and $\omega_{2}$ via $R_4$. The beams are applied for $100\,\mu\text{s}$, with an estimated Rabi frequency of $\Omega_R \approx 2\pi \times 20$\,kHz.

\subsection{Cooling Sequences}
Prior to Raman sideband cooling, we optically pump molecules into $\left|\uparrow\ket$ by applying beam OP for 2\,ms.

For the radial/axial cooling sequence (RAC), both radial beams $R_1$ and $R_2$ are on and retro-reflected. Axially, we send $\omega_{1}$ into $R_3$, and $\omega_{2}$ into $R_4$. We perform Raman transfer on the radial $\Delta n_r = -2$ sideband for $150\,\mu\text{s}$, and optical pumping with OP for $150\,\mu\text{s}$. The Raman transfer occurs at tweezer depth $U_R$, and the optical pumping occurs at depth $U_0$. The tweezer depth ramps occur over $200\,\mu\text{s}$ between each of these steps.

For the interlaced radial cooling sequence (IRC), we only turn on radial beams $R_1$ and $R_2$, with $R_2$ retro-reflected. Raman transfer on the $\Delta n_r = -2$ ($\Delta n_r=-1$) sideband occurs for $150\,\mu\text{s}$ ($300\,\mu\text{s}$).

\bibliographystyle{apsrev4-1}

\end{document}


\title{Supplementary Information for ``Raman Sideband Cooling of Molecules in an Optical Tweezer Array" }
\author{Yukai Lu}
\thanks{These authors contributed equally to this work.}
\affiliation{Department of Physics, Princeton University, Princeton, New Jersey 08544 USA}
\affiliation{Department of Electrical and Computer Engineering, Princeton University, Princeton, New Jersey 08544 USA}
\author{Samuel J. Li}
\thanks{These authors contributed equally to this work.}
\affiliation{Department of Physics, Princeton University, Princeton, New Jersey 08544 USA}
\author{Connor M. Holland}
\thanks{These authors contributed equally to this work.}
\affiliation{Department of Physics, Princeton University, Princeton, New Jersey 08544 USA}
\author{Lawrence W. Cheuk}
\email{lcheuk@princeton.edu}
\affiliation{Department of Physics, Princeton University, Princeton, New Jersey 08544 USA}

\maketitle

Included in this document is supplementary information for: 1) bright state admixture due to trapping light, 2) heating during optical pumping, 3) Raman sideband broadening mechanisms and estimates, 4) Raman thermometry, and 5) relation of entropy to temperature.

\section{Bright State Admixture due to Trapping Light}
\label{sm:optical-pumping}
In this section, we discuss the bright state admixture into the stretched states $\ket{\pm} =\ket{ F=2, m_F=\pm2} $ in the $X(v=0,N=1)$ manifold, relevant to the discussion of optical pumping in the main text. 
The states $\{\ket{+},\ket{-}\}$ can be prepared via optical pumping as each is dark to $\{\sigma_+, \sigma_-\}$ and $\pi$ light addressing the $X^2\Sigma(v=0, N=1)-A^2\Pi_{1/2} (v=0, J=1/2, +)$ transition. For example, one can optically pump into $\ket{+}$ by applying light devoid of a $\sigma_-$ component.

In the presence of tweezer light, the free space eigenstates $\ket{\pm} =\ket{ F=2, m_F=\pm2} $ are perturbed and become slightly modified eigenstates $\ket{\widetilde{\pm}}$ containing other free-space eigenstates that are bright to $\sigma_{\pm}$ or $\pi$ light. For example, the in-trap eigenstate $\ket{\widetilde{+}}$ can be decomposed into free-space eigenstates as $\ket{\widetilde{+}}=\ket{+}+\epsilon_i\ket{+_{i}}$, where $\ket{+_{i}}$ are other $N=1$ states that are bright to $\sigma_+$ or $\pi$ light. The bright state population admixture $\mathcal{M}=\sum_i |\epsilon_i|^2$ therefore quantifies the brightness of $\ket{\widetilde{+}}$ to optical pumping light, which determines the quality of optical pumping.

\begin{figure}[h!]
	\includegraphics[width=\columnwidth]{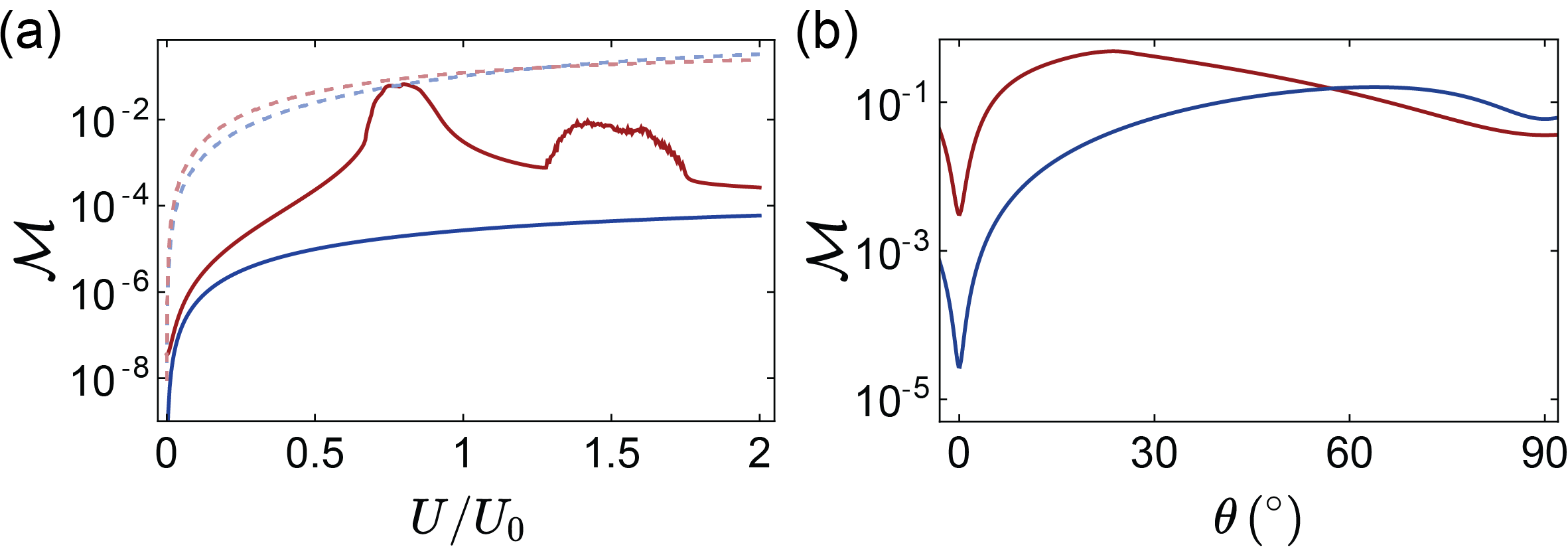}
	\caption{Bright State Admixtures. (a) The admixtures $\imp{-}$ (blue) and $\imp{+}$ (red) of $\ket{\widetilde{\pm}}$ versus trap depth $U$, at $\theta=0^\circ$ and a magnetic field of $B=\SI{4.4}{G}$ (solid), and $\theta=\theta_m$ and $B = \SI{5.5}{G}$ (dashed). The curves are computed for a thermal ensemble with reduced temperatures $(\tilde{T}_r, \tilde{T}_z) = (2.5, 12)$ where $\tilde{T}_\alpha=k_BT/\hbar\omega_\alpha$, $\omega_r (\omega_z)$ being the radial (axial) trapping frequency. 
 (b) The admixtures $\imp{-}$ (blue) and $\imp{+}$ (red) versus angle $\theta$ at a tweezer depth of $U_0 = k_B \times 930\,\mu\text{K}$ and $B = \SI{5.5}{G}$.}
	\label{fig:pumping-darkness}
\end{figure}

To determine whether $\ket{\widetilde{+}}$ or $\ket{\widetilde{-}}$ is better for optical pumping, we numerically compute $\mathcal{M}$ in the presence of optical tweezer light at 781\,nm. Specifically, we compute $\mathcal{M}$ via $\mathcal{M}_{\pm} = 1- \left|\langle \pm |\widetilde{\pm} \rangle\right|^2$ both as a function of the tweezer depth ($U$) and as a function of the angle ($\theta$) between the bias magnetic field and the tweezer polarization axis. As shown in Fig. \ref{fig:pumping-darkness}, the darkness is maximal at $00^\circ$, and $\ket{\widetilde{-}}$ is darker than $\ket{\widetilde{+}}$ across all relevant depths.

To elaborate, while $\ket{\widetilde{\pm}}$ are equivalent choices at zero magnetic field, in the presence of a magnetic field and optical trapping light, the two behave differently. The tensor ac Stark shifts arising from the tweezer light lower the energies of the stretched states $\ket{\widetilde{\pm}}$ relative to the other $F=2$ states. These shifts, along with Zeeman shifts in the presence of a magnetic field, leave $\ket{\widetilde{-}}$ lower in energy than all other $F=2$ states, minimizing level crossings and decreasing its bright state population admixture. We therefore choose to optically pumping molecules into $\ket{\widetilde{-}}$ using light devoid of a $\sigma_+$ component. In the main text, this state is denoted as $\ket{\uparrow}$. Fig. \ref{fig:pumping-darkness}(b) illustrates how the darkness is greatly reduced when the $B$-field is not parallel to the tweezer polarization. This arises due to larger $\sigma_+$ and $\sigma_-$ components in the tweezer light when the quantization axis is away from $\theta=0^\circ$, leading to increased coupling to bright states. This limits the achievable darkness at the magic angle $\theta = \theta_m$ for moderate tweezer depths relevant for Raman sideband cooling.

\section{Heating During Optical Pumping }

\begin{figure}[h!]
	\includegraphics[width=\columnwidth]{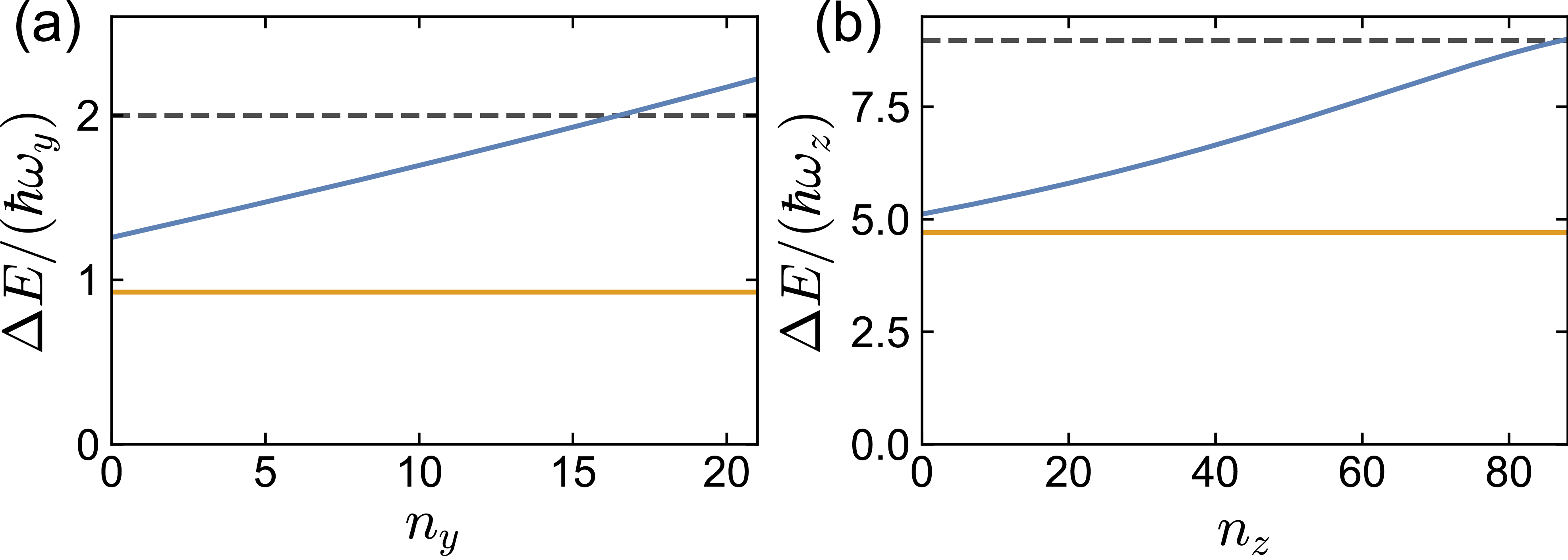}
        \caption{Motional energy increase during optical pumping. The computed energy increase is shown (a) versus radial motional quantum number $n_y$ along $\hat{y}$, with $n_x=0$ and $\bar{n}_z = 26$, and (b) versus axial motional quantum number $n_z$ with $\bar{n}_x=\bar{n}_y=1.4$. Blue (orange) curves do (do not) include projection heating, computed similarly as described in~\cite{caldwell2020sideband}. The dashed lines indicate the energy removed during radial/axial cooling (RAC) that addressing the $\Delta n_r = -2$ sideband.}
	\label{fig:opheat}
\end{figure}

In free-space, each photon emitted or absorbed will increase the motional energy by an amount $E_R = \hbar^2 k^2/2m = h \times 9.2$\,kHz. In our optical pumping configuration, an average of 19 total photons are required to pump the molecule from $\ket{\downarrow}$ to $\ket{\uparrow}$. In terms of radial motional quanta, this leads to a heating of $\Delta n_r \sim 19 E_R / (\hbar \omega_r) = 0.93$. In addition to this recoil heating, projection heating occurs as the molecule transitions between the various $N=1$ hyperfine states in the presence of differing trapping potentials~\cite{caldwell2020sideband}. Projection heating worsens with increasing motional quantum number (Fig.~\ref{fig:opheat}). Starting from our initial $\tilde{T}_r = 1.8$ and $\tilde{T}_z = 26.5$, projection heating contributes an additional $0.40 \hbar \omega_r$ of energy each cooling cycle. To obtain net cooling, it is therefore necessary to address the $\Delta n = -2$ radial sideband. At this two-photon Raman detuning, the corresponding vertical Raman beams address the $\Delta n \approx -9$ axial sideband.

\section{Raman Sideband Broadening Mechanisms and Estimates}
In this section, we describe and estimate various mechanisms that broaden Raman spectra.

In an ideal scenario, the internal states $\ket{\uparrow}$ and $\ket{\downarrow}$ experience identical harmonic potentials with trapping frequency $\omega$, giving rise to motional states separated in energy by $\hbar\omega$. Motional changing transitions of the form $\ket{\uparrow,n_i} \to \ket{\uparrow,n_f = n_i + \Delta n} $ can be driven by addressing Raman sidebands using a two-photon detuning $\delta$ of
\begin{equation}
    \hbar \, \delta(\Delta n)
  = E_0 + \hbar \omega \Delta n,
\end{equation}
where $E_0$ is the carrier $(\Delta n=0)$ energy separation. In this idealized approximation, motional changing transitions are grouped into sidebands which are spaced by multiples of $\hbar\omega$ about the carrier.

\subsection{Trap Anharmonicity and Differential Trapping}
Trapping anharmonicity, which arises from the Gaussian trapping profile of the tweezer beams, and state-dependent ac Stark shifts, which arises from the complex internal structure of molecules, lead to departures from this ideal scenario. In particular, trapping anharmonicity leads to motional state-dependent sideband spacing, while differential ac Stark shifts lead to motional state-dependent carrier energies and the sideband spacings. At finite motional temperature, these state-dependent effects lead to broadening of both the carrier and sidebands.

Since these mechanisms affect the carrier energy and sideband spacings differently, we parametrize the transition energies as
\begin{equation}\label{eq:transition freq}
    \hbar \, \delta(\Delta n, n_i,n_y,n_z)    = E_0(n_i, n_y, n_z) + \hbar \omega(\Delta n, n_i) \, \Delta n,
\end{equation}
where we assume we are driving motion-changing transitions along $\hat{x}$ in a 3D trap. 

The carrier energy can then be expanded to low order as 
\begin{align*}
    E_0(n, n_y, n_z) = E_0(0) &+ \hbar \omega_0 \, (\epsilon_1 n + \epsilon_2 n^2 + \cdots \\
                             &+\epsilon_{1y} n_y + \epsilon_{2y}n_y^2 + \cdots \\
                             &+\epsilon_{1z} n_z + \epsilon_{2z} n_z^2 + \cdots),
\end{align*}
and the sideband spacing as
\begin{equation}
    \omega(\Delta n, n) = \omega_0 \, (1 + \kappa_{1,0} n + \kappa_{0,1} \Delta n + \kappa_{2, 0} n^2 + \cdots)
\end{equation}
in terms of dimensionless coefficients $\kappa, \epsilon$.

The relative broadening of the $\Delta n$ sideband compared to the trapping frequency $\omega_0$ is therefore determined by the variation of
\begin{equation}
\delta( \Delta n,n_i)/\omega_0
    = (\epsilon_1 + \kappa_{1,0} \Delta n) n_i + \epsilon_{1y} n_y + \epsilon_{1z} n_z + \cdots
\end{equation}
over the thermal distribution of $n_i, n_y, n_z$. When this is $\leq 1$, the sideband is resolved.

\begin{figure}
	\includegraphics[width=\columnwidth]{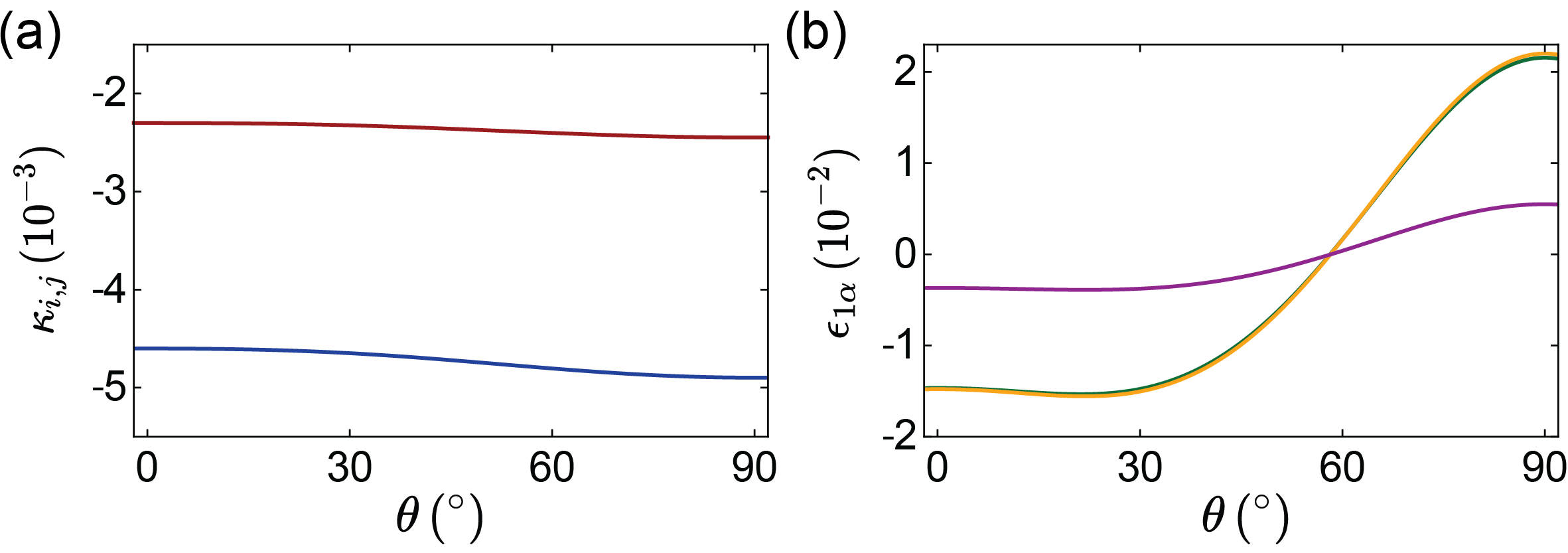}
	\vspace{-0.25in}
	\caption{Dimensionless expansion coefficients versus angle $\theta$. We fix $U=U_R$ and $B = \SI{5.5}{G}$. (a)  $\kappa_{1, 0}$($\kappa_{0,1}$) versus $\theta$ (blue, red). (b) $\epsilon_{1}$ (green) , $\epsilon_{1y}$ (orange), and $\epsilon_{1z}$ (purple) are shown versus $\theta$.}
	\label{fig:broadening-theory}
\end{figure}

\begin{figure}
    \includegraphics[width=\columnwidth]{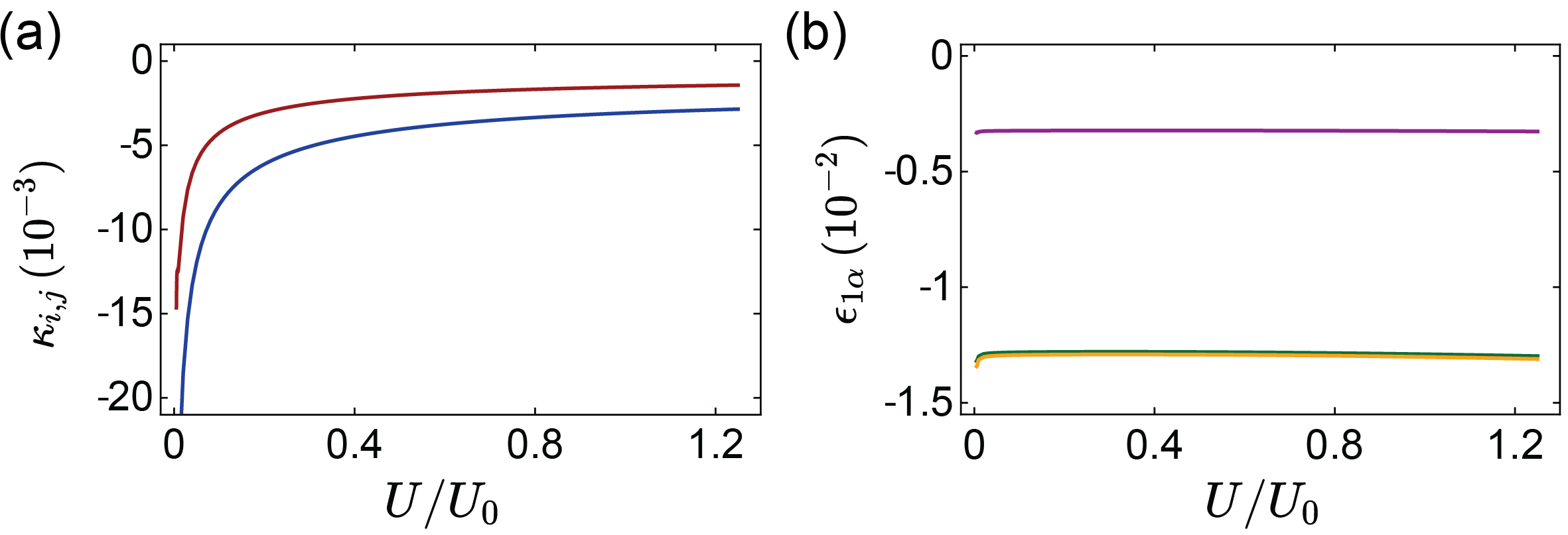}
    \vspace{-0.25in}
    \caption{Dimensionless expansion coefficients as a function of trap depth $U$.
        We fix $B = \SI{4.4}{G}$ and $\theta = 0^\circ$. (a)$\kappa_{1, 0}$ ($\kappa_{0,1}$) (blue, red) versus $U$.  (b) $\epsilon_{1}$ (green), $\epsilon_{1y}$ (orange), and $\epsilon_{1z}$ (purple) versus $U$.}
    \label{fig:broadening-theory-depth}
\end{figure}

In Figures~\ref{fig:broadening-theory} and \ref{fig:broadening-theory-depth}, we show the expansion coefficients $\kappa_{0,1}, \kappa_{1,0}, \epsilon_{1\alpha}$ computed for our experimental parameters using a full model of the tweezer light field. We find that higher order expansion coefficients are negligible. Near $\theta\approx\SI{57}{\degree}$, the parameters $\epsilon_{1\alpha}$ vanish nearly simultaneously, strongly narrowing the carrier ($\Delta n=0$) transition, as described in the main text.

\subsection{Tweezer Depth Inhomogeneity}
In addition to the trap anharmonicity and differential ac Stark shifts, the Raman transitions are further broadened by tweezer depth inhomogeneity. We estimate that the tweezer trap depth $U$ varies fractionally by $\widetilde{\delta U} = \delta U/U \sim 0.05$ across the 37-site tweezer array.
This inhomogeneity leads to variations in the carrier transition $E_0(0)$ and the trapping frequency $\omega_0$, contributing an additional broadening of
\begin{equation}
    \frac{U}{\hbar} \dv{E_0(0)}{U} \widetilde{\delta U} + \Delta n \, U \dv{\omega_0}{U} \widetilde{\delta U}
\end{equation}
to the $\Delta n$ sideband. We note that the effect of tweezer depth inhomogeneity can be suppressed by minimizing differential ac Stark shifts.

\subsection{Accounting of Raman Broadening}
In Table~\ref{tab:broadening}, we summarize the various broadening contributions to the  $(\Delta n=0)$ carrier linewidth both in the cooling ($\theta=0^\circ$) and magic ($\theta=\theta_m$) configurations.
The calculation is done using reduced temperatures $(\tilde{T}_x, \tilde{T}_y, \tilde{T}_z) = (2.5, 2.5, 12)$ prior to Raman sideband cooling, where $\tilde{T} = k_B T / \hbar \omega$. 
With the exception of Fourier broadening due to the finite Raman pulse time, a dominant broadening mechanism appears to be tweezer inhomogeneity. The experimental linewidth is therefore a sensitive probe of the differential ac Stark shift between the Raman states.

\section{Resolved Axial Sidebands}

\begin{figure}
	\includegraphics[width=\columnwidth]{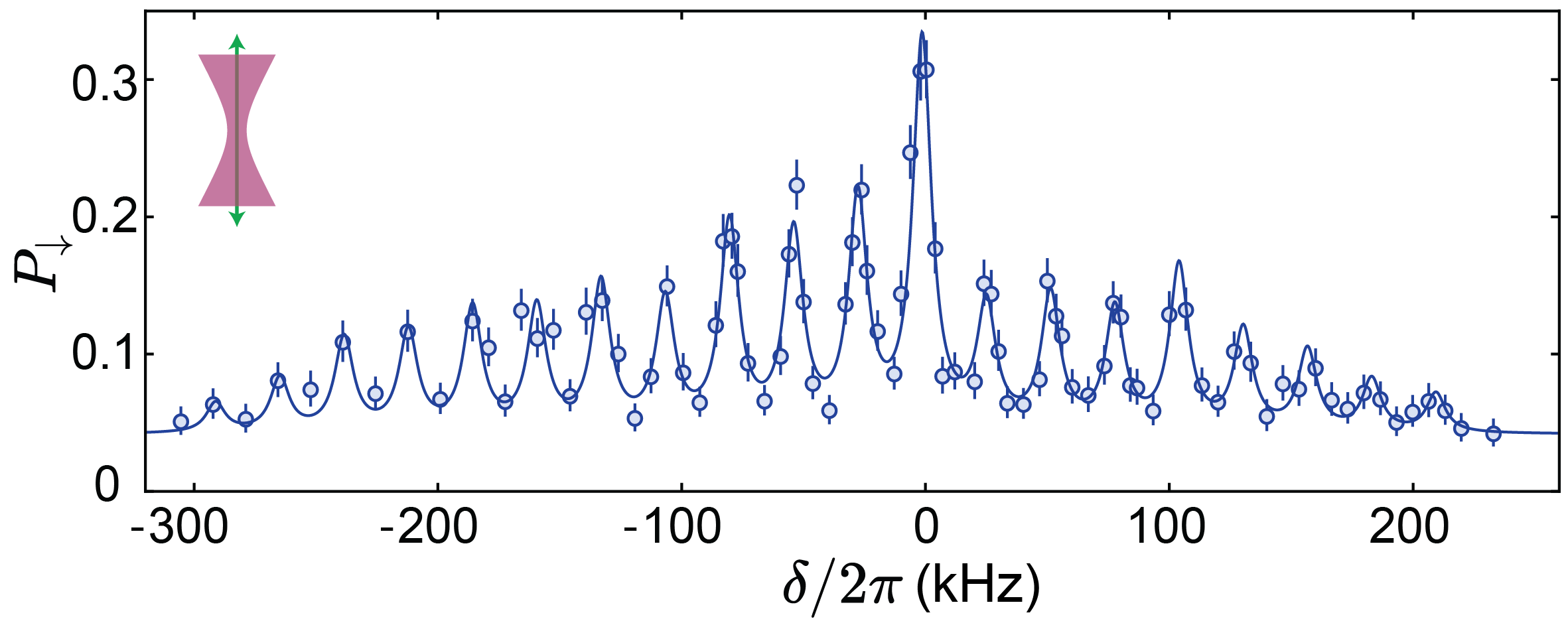}
	\vspace{-0.25in}
	\caption{Resolved Axial Spectrum at Magic Configuration. Blue circles show transfer ($P_\downarrow$) following interlaced radial cooling (IRC) as a function of detuning $\delta$ from the carrier. Solid blue curve is a fit to 20 equally spaced Lorentzians with independent heights and an offset. Sidebands are visible between $\Delta n_z=-8$ and $\Delta n_z=11$.}
	\label{fig:ResolvedAxSpect}
\end{figure}

The Raman linewidth in the magic configuration ($\Gamma \approx 2\pi\times7\,\text{kHz}$) is lower than the axial trapping frequency ($\omega_z \approx 2\pi \times26\,\text{kHz}$). This, combined with the relatively large axial Lamb-Dicke parameter $\eta_z=1.34$ and high $\bar{n}_z$, enables resolving sidebands up to $|\Delta n_z|\sim10$.  Fig.~\ref{fig:ResolvedAxSpect} shows the resolved axial Raman spectrum taken in the magic configuration.

\begin{table}[b]
	\begin{tabular}{l|r|r}
		Contribution & Cooling (kHz) & Magic (kHz) \\
		\hline
		Fourier Limit & 5.2 & 5.2 \\
		$\omega_0 \epsilon_1 \, \delta n_i$ & 6.9 & $<0.1$ \\
		$\omega_0 \epsilon_{1y} \, \delta n_y$ & 7.0 & $<0.1$ \\
		$\omega_0 \epsilon_{1z} \, \delta n_z$ & 8.5 & $<0.1$ \\
		Depth Inhomogeneity & 23.2 & 1.5 \\
		$\omega_0 \epsilon_2 \, \delta(n_i^2)$ & $<0.1$ & $<0.1$ \\
		$\omega_0 \epsilon_{2y} \, \delta((n_y)^2)$ & $<0.1$ & $0.1$ \\
		$\omega_0 \epsilon_{2z} \, \delta((n_z)^2)$ & 0.1 & $0.4$ \\
		\hline
		Total & 27.1 & 5.4 \\
		Experiment & 26.5(3) & 7.1(2)
	\end{tabular}
	\caption{Contributions to the carrier linewidth in the cooling configuration ($B = \SI{4.4}{G}$, $\theta = 0^\circ$) and the magic configuration used for Raman spectroscopy ($B = \SI{5.5}{G}$, $\theta = \theta_m$). Experimental linewidths are taken with \SI{170}{\micro\second} interrogation time. Linewidths are full-width at half-max values, and the total is the quadrature sum of the individual components.}
	\label{tab:broadening}
\end{table}

\section{Raman Thermometry}
In this section, we describe Raman thermometry in both the resolved and unresolved limits. As described in the main text, the motional sidebands are resolved for the radial direction, while less so for the axial direction. As we will show, robust and quantitative thermometry can be performed in both limits.

\subsection{Resolved Sideband Thermometry along the Radial Direction}
\begin{figure}
	\includegraphics[width=\columnwidth]{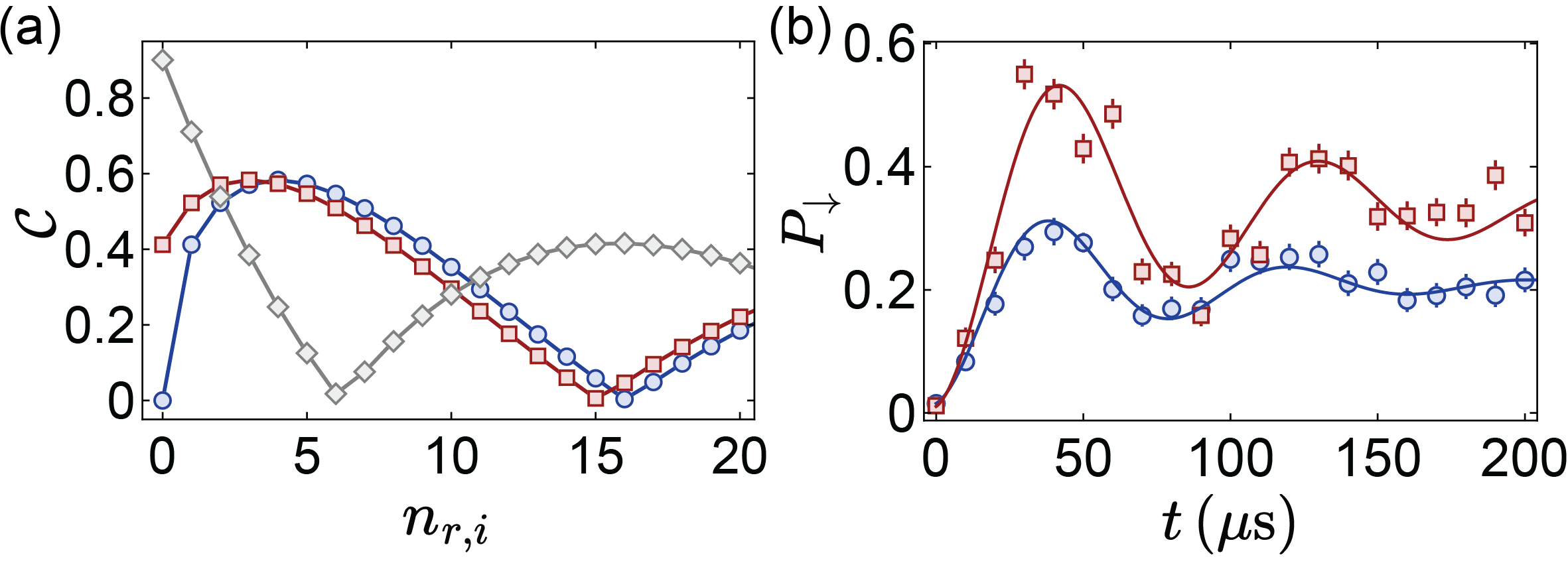}
	\vspace{-0.25in}
	\caption{Motional Transfer Dynamics and Coupling Strengths: (a) Motional coupling strength $\mathcal{C}$ as a function of initial radial state $n_{r,i}$ for $\Delta n_r=-1$ (blue circles), $\Delta n_r=0$ (gray diamonds), and $\Delta n_r=+1$ (red squares). (b) Blue circles (red squares) show population transferred to $|{\downarrow}\rangle$ as a function of Raman transfer time using the $\Delta n_r = -1\, (+1)$ sideband, demonstrating coherent dynamics. Solid curves are fits to damped sinusoids.}
	\label{fig:CohOsc}
\end{figure}

To measure the ground state fraction and radial temperature, we probe Raman transfer on the $\Delta n= \pm1$ sidebands. In our temperature ($\bar{n}_r<1$) and trapping ($\eta_r=0.46$) regimes during Raman transfer, the motional Raman coupling is approximately constant for all significantly occupied motional states (Fig.~\ref{fig:CohOsc}(a)). This allows us to observe coherent Rabi oscillations on both $\Delta n= \pm1$ sidebands (Fig.~\ref{fig:CohOsc}(b)).
 
To perform radial thermometry, we apply a $\pi$-pulse to probe population transferred by addressing the $\Delta n= \pm1$ sidebands. The resulting spectra are simultaneously fit to Lorentzians with a common trapping frequency and a background offset. The fitted offset agrees with the independently measured value. The peak heights $M_{\pm 1}$ can be used to extract both the 1D radial ground state fraction and the radial temperature. Because the motional coupling is approximately uniform, the 1D radial ground state fraction $P_0$ can be obtained from $M_{\pm 1}$ via
\begin{equation}
P_{0} = \frac{M_+-M_-}{M_+} = 1- \frac{M_-}{M_+} = 1-\frac{1}{\mathcal{A}},
\end{equation}
where we define the asymmetry $\mathcal{A}=M_{+}/M_{-}$. This expression is also valid for non-uniform couplings, as long as the motional states are thermally occupied. 

Assuming a thermal occupation of radial motional states, we find $\mathcal{A}=e^{\hbar\omega_r/(k_B T_r)}$, which can be inverted to give 
\begin{equation}
k_B T_r = \frac{\hbar \omega_r}{ \ln \mathcal{A}}
\end{equation}
The average motional quanta is also related to the asymmetry through $\mathcal{A} = (\bar{n}_r+1)/\bar{n}_r$. Inverting this relation gives
\begin{equation}
 \bar{n}_r =  \frac{1}{\mathcal{A}-1}
\end{equation} 

\subsection{Two-Photon Doppler Thermometry in Free Space}
To guide the discussion of Raman thermometry in the high-temperature and unresolved limit, we first describe thermometry in free space using a two-photon transition. We will refer to this as two-photon Doppler thermometry.

In free space, momentum, parameterized by a wavevector $k$, is a good quantum number, A Raman process connects $\ket{\uparrow, k}$ to $\ket{\downarrow, k+ \Delta k}$, where $\Delta k$ is the difference in the wavevectors between the two Raman beams. The Hamiltonian can therefore be written as
\begin{eqnarray}
\hat{H} &=& \sum_k \hat{H}_k \nonumber \\
\hat{H}_k &=& \left(\begin{array}{cc} 
 \frac{\hbar^2}{2m}  k^2 & \frac{\hbar \Omega_R}{2} \\
 \frac{\hbar \Omega_R}{2}  & \frac{\hbar^2}{2m} (k+\Delta k)^2
\end{array}\right) \nonumber \\
& =& \frac{\hbar^2 k^2}{2m} \mathbf{1} +\frac{\hbar\Omega_R}{2} \sigma_x +\left(\begin{array}{cc} 
0 & 0 \\
0& \frac{\hbar^2}{m} (k\Delta k) + E_R
\end{array}\right) ,
\end{eqnarray}
where $E_R= \frac{\hbar^2}{2m} (\Delta k)^2$ is the recoil energy associated with $\Delta k$. Therefore, relative to motion-insensitive Raman transfer with copropagating Raman beams, the two-photon resonance shifts by $ (\Delta k v) + E_R/\hbar$, where $v=\hbar k/m$. In other words, a two-photon Doppler shift and recoil shift are present. For a thermal distribution where the population follows $P(v) \propto \exp(- (mv^2)/(2k_B T))$, the resulting Raman spectrum is Gaussian. In the short time limit, population transfer is proportional to the population in each velocity class, and hence is given by
\begin{equation}
P(\delta)  \propto \exp\left(-\frac{m}{2k_B T}\left(\frac{\hbar \delta-E_R}{\hbar \Delta k}\right)^2\right).
\label{eq:doppler}
\end{equation}

\subsection{Unresolved Thermometry in the Doppler Limit along the Axial Direction}
Since the axial trapping frequencies are smaller than the radial ones, the axial sidebands are more closely spaced and are challenging to resolve. Furthermore, because the axial Lamb-Dicke parameter is large, there is significant and non-uniform coupling to many sidebands, complicating sideband thermometry. Specifically, extracting the ground state fraction and determining axial temperatures using sideband asymmetry are challenging.

To perform robust axial thermometry, we instead measure Raman spectra far from the carrier and in the unresolved regime. Specifically, we increase the Raman Rabi coupling until high-order sidebands become unresolved and blend into the Gaussian-shaped wings. The resulting wings are fit to a Gaussian, and an axial temperature is extracted by comparing the fit to a numerical simulation. This method is agnostic to many details of the axial trapping potential and is equivalent to two-photon Doppler spectroscopy.

\begin{figure}
	\includegraphics[width=\columnwidth]{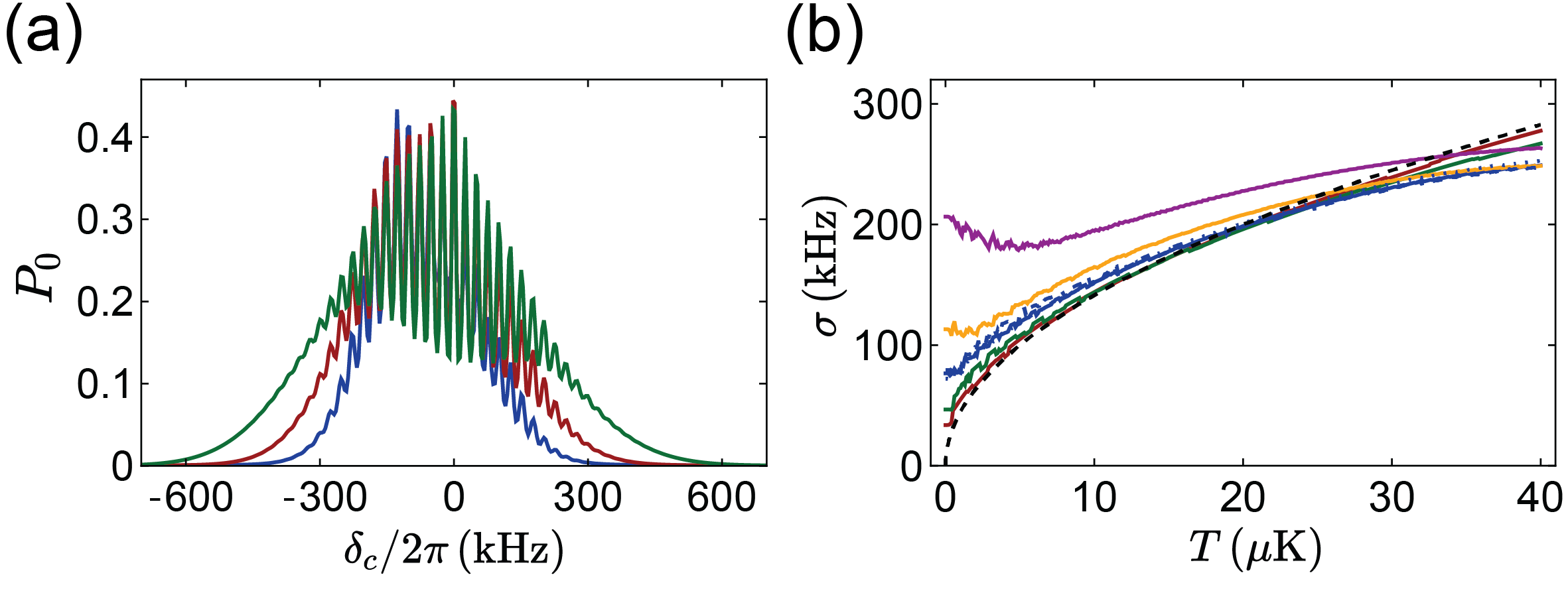}
	\caption{Simulation of Axial Spectra and Spectral Widths. (a) Simulated spectra at \SI{5}{\mu\text{K}} (blue), \SI{10}{\mu\text{K}} (red), and \SI{20}{\mu\text{K}} (green), using the experimentally determined Rabi frequency $\Omega_R = 2\pi \times 20(2)$ kHz and transfer time $\tau = \SI{100}{\mu\text{s}}$. (b) Gaussian widths (fitted to the wings) versus temperature $T$, for Rabi frequencies $\Omega_R = 2\pi \times \SI{1}{kHz}$ (red), \SI{10}{kHz} (green), \SI{20}{kHz} (blue), \SI{30}{kHz} (orange), and \SI{50}{kHz} (purple). Solid curves are for a tweezer inhomogeneity of $\widetilde{\delta U} = 5\%$. The broken blue curves correspond to $\Omega_R = 2\pi \times \SI{20}{kHz}$ with $\widetilde{\delta U} = 1\%$ (dashed) and $10\%$ (dotted). The black dashed line shows the analytical two-photon Doppler result.}
	\label{fig:simulation}
\end{figure}

\subsubsection{Simulation of Axial Sideband Spectrum}
Axial temperatures reported in the main text are obtained by fitting the measured axial Raman spectra to a full numerical simulation of the Raman transfer dynamics. Our simulations confirm that radial motion has negligible effect on the axial spectrum, since differential ac Stark shifts are nearly zero in the magic configuration. Consequently, we ignore radial motion in the following discussion.

In detail, the trapping light field (including polarization) is numerically computed using the separately characterized beam waist and depth~\cite{holland2022bichromatic}. At each point, we diagonalize the effective Hamiltonian $H_{\text{eff}} = H_{0} + H_{t}$ for the $X^2\Sigma(v=0, N=1)$ rovibrational manifold, where $H_0$ is the free-space molecular Hamiltonian in the presence of a magnetic field, and $H_t$ describes the effect of the tweezer light. This yields effective potentials for the different internal states. We compute the lowest $n_{\text{mot}} = 512$ motional eigenstates $\ket{\psi_i}$ in each of these potentials corresponding to the 12 $X^2\Sigma(v=0, N=1)$ hyperfine states.

To simulate transfer dynamics, we make the rotating wave approximation, and describe the system with the Hamiltonian
\begin{equation}
    H_{\text{mot,eff}} = \mqty[H_{\uparrow} + \delta \mathbf{1}& V_{I} \\ V_{I}^\dagger & H_{\downarrow}],
     \end{equation}
where $V_{I}$ encodes Raman coupling between motional states, $H_{\uparrow}$ ($H_{\downarrow}$) are diagonal matrices containing the energies of the motional eigenstates for $\ket{\uparrow}$ ($\ket{\downarrow}$) molecules, and $\delta$ is the two-photon detuning. $V_{I}$ has elements
\[
   (V_{I})_{m,n} = \frac{\hbar \Omega_R}{2} \mel{\psi_m}{e^{i \Delta k_z z}}{\psi_n},
\]
where $\Omega_R $ is the two-photon Rabi frequency.

To model broadening due to trap depth inhomogeneity, we perform the simulations over a distribution of trap depths with $\widetilde{\delta U} = 5\%$ RMS variation, and average the results. 

We find that the simulated axial spectrum displays Gaussian-like tails (Fig.~\ref{fig:simulation}(a)), whose fitted Gaussian widths increase monotonically for sufficiently low $\Omega_R$. At low $\Omega_R$, 
the widths approach that of two-photon Doppler spectroscopy in free space (Fig.~\ref{fig:simulation}(b)), a fact we justify in the high-temperature limit ($n_z\gg 1$). At much larger $\Omega_R$, saturation effects cause substantial deviations in the shape of the spectrum, rending the fitted Gaussian width unreliable.

To extract an axial temperature, we fit the tails of our measured axial spectra to Gaussians. The fit widths are converted to temperatures via numerical simulations incorporating our independently estimated $\Omega_R=2\pi \times 20$\,kHz. Simulations show that the fitted widths are monotonic with temperature, robust to trap inhomogeneity, and similar to those obtained in the free-space two-photon Doppler spectroscopy limit given our Rabi frequency and axial temperature regime ($T_z\sim 5 - 40\,\mu\text{K}$).

\subsubsection{Equivalence of Raman Spectroscopy with Two-Photon Doppler Spectroscopy in the High-Temperature Limit}
\label{subsec:doppler}
In this section, we show that Raman spectra reduce to two-photon Doppler spectra in certain limits. Physically, this occurs when the Raman transfer is between trapped eigenstates that sufficiently resemble free-space motional states. As we will show, the two become equivalent under the following conditions:
\begin{enumerate}
\item The motional states are thermally populated.
\item Saturation is negligible (i.e. Raman transfer is in the short time unsaturated limit ).
\item The temperature is sufficiently high compared to the trapping frequency, that is, the reduced temperature $\tilde{T} = k_B T / \hbar \omega$ is $ \gg 1$.
\item In the case of tight confinement, the reduced temperature satisfies $\eta^2 \tilde{T} \gg 1$, where $\eta$ is the Lamb-Dicke parameter.
\end{enumerate}
The Lamb-Dicke parameter is defined as $\eta = \Delta k\, l/\sqrt{2}$, where $l = \sqrt{\frac{\hbar}{m \omega}}$ is the harmonic oscillator length. $\eta$ can also be expressed in terms of $\omega$ and $E_R$ via $\eta = \sqrt{E_R/\hbar \omega}$.

We assume that the trapping potentials are identical for both Raman states, and the trapping is harmonic. The motional coupling matrix element is then given by~\cite{Wineland1979lasercooling}
\begin{equation}
    \mel{n}{e^{i \Delta k \, x}}{n'} = e^{-\eta^2/2} \sqrt{\frac{n_<!}{n_>!}} \eta^{\abs{n - n'}} L_{n_<}^{\abs{n - n'}}(\eta^2)
\end{equation}
for the motional coupling amplitudes, where $n_<, n_> \defeq \min(n, n'), \max(n, n')$, and $L_n^\alpha$ is the generalized Laguerre polynomial.

We next consider the population transfer on the $\Delta n = q$ sideband in the short time limit. Assuming a thermal distribution, we apply Fermi's golden rule and find that the transferred population $P(q)$ follows
\begin{equation}
    P(m) \propto \lambda (q, \tilde{\beta}, \eta) \defeq  \frac{1}{Z} \sum_{n=\text{max}(0,-q)}^{\infty} e^{-\tilde{\beta} n}\left|\mel{n}{e^{i \Delta k \, x}}{n'}\right|^2, \nonumber
\end{equation}
where $\tilde{\beta} = 1/\tilde{T} = (\hbar \omega)/(k_B T)$ and  $Z=\sum_{n=0}^{\infty} e^{-\tilde{\beta}n}$ is the partition function. 

We first treat the case for $q>0$. Explicitly, $\lambda (q, \tilde{\beta}, \eta)$ is given by
\begin{equation}
\lambda(q, \tilde{T}, \eta) =
    e^{-\eta^2} \sum_{n = 0}^\infty \frac{e^{-\tilde{\beta} n}}{Z} \eta^{2m} \frac{n!}{(n+q)!} L_{n}^q(\eta^2)^2. \nonumber
\end{equation}
Using the Hardy-Hille formula \footnote{W.A.Al-Salam 1964}
\begin{equation}
    \sum_{n=0}^\infty \frac{n! \, t^n \, L_n^\alpha(r) L_n^\alpha(s) }{\Gamma(n+\alpha+1)}
    = \frac{e^{-(r+s)t/(1-t)}}{(rst)^{\alpha/2} (1-t)} I_\alpha\left(\frac{2\sqrt{rst}}{1-t}\right) \nonumber
\end{equation}
with $r=s=\eta^2$ and $t = e^{-\tilde{\beta}}$, $\lambda(q, \tilde{\beta}, \eta)$ can be written as
\begin{equation}
    \lambda(q, \tilde{\beta}, \eta) = e^{-\eta^2} t^{-q/2} e^{-2\eta^2 t/(1-t)} I_q\left(\frac{2 \eta^2 \sqrt{t}}{1-t}\right) \nonumber
\end{equation}
where $I_q(x)$ is the $q$th modified Bessel function of the first kind.

In the high temperature limit where $\tilde{\beta}\ll 1$ (assumption 3), $t\approx 1-\beta$. In this limit, the argument of the Bessel function, $x= 2 \eta^2 \sqrt{t}/(1-t)$, approaches $2 \eta^2 \tilde{T}$. Assuming that the reduced temperature $\tilde{T}$ is further much larger than $1/\eta^2$ (assumption 4), we can use an asymptotic expansion for $I_m(x)$ in the $x\gg 1$ limit to obtain ~ \footnote{Frolich and Spencer 1981 Appendix B}

\begin{eqnarray}
    \lambda(q, \tilde{T}, \eta)& \sim& \frac{e^{-x \sqrt{t} - \eta^2}}{\sqrt{2\pi x} \sqrt[4]{1 + (q/x)^2}} \exp(g(q, x)), \nonumber \\
    g(q,x) &=& q \frac{\tilde{\beta}}{2} - q \, \mathrm{arsinh}\left(\frac{q}{x}\right) + \sqrt{x^2 + q^2}. \nonumber
\end{eqnarray}
For fixed $x$, the function $g(q,x)$ has a unique maximum at $q=\eta^2$. Expanding to second order in $q$ about the maximum, we obtain

\begin{equation}
\lambda(q, \tilde{T}, \eta) \sim \frac{e^{-x\sqrt{t} - \eta^2 + g(\eta^2,x)}}{\sqrt{2\pi x} \sqrt[4]{1 + (q/x)^2}} \exp(-\frac{1}{2} \left(\frac{q - \eta^2}{\sigma}\right)^2),
\end{equation}
where $\sigma =\eta \sqrt{\frac{1+t}{1-t}}$. For $\tilde{T} \gg 1$, we have $\sigma \sim \eta \sqrt{2 \tilde{T}}$. Near the maximum, the prefactor for the Gaussian function is approximately constant. The $q$-dependence of the prefactor arises from the factor $A(q) =1/\sqrt{1+(q/x)^2} \approx 1/\sqrt{1+(q/(2\eta^2\tilde{T}))^2}$. The Gaussian envelop is only significant around a range in $q$ of $\sigma \sim \eta \sqrt{2\tilde{T}}$ about $q=\eta^2 \ll \eta^2 \tilde{T}$. Therefore, when $\eta^2 \tilde{T}\gg 1$ (assumption 4), $A(q)$ only varies by $1/(\eta^2 \tilde{T})$ over the relevant range, and is approximately constant. Thus for $q>0$,
\begin{equation}
P(q) \propto \exp(-\frac{1}{2} \left(\frac{q - \eta^2}{\sigma}\right)^2) \nonumber 
\end{equation}
The argument can be extended for $q<0$ to give
\begin{eqnarray}
P(q)&\propto & e^{-\tilde{\beta} \abs{q}} \exp(-\frac{1}{2} \left(\frac{|q| - \eta^2}{\sigma}\right)^2) \nonumber \\
&=& \exp(-\frac{1}{2} \left(\frac{q - \eta^2}{\sigma}\right)^2). \nonumber
\end{eqnarray}
Therefore, under assumptions (1-4), the Raman transferred population approaches the following form:
\begin{equation}
P(q) \propto  \exp(-\frac{1}{2} \left(\frac{q - \eta^2}{\sigma}\right)^2). \nonumber 
\end{equation}
In terms of the two-photon detuning $\delta$, one can restore dimensions to obtain
\begin{eqnarray}
P(\delta) &\propto& \exp(-\frac{1}{2}\frac{ \left(\delta/\omega- E_R/(\hbar \omega)\right)^2}{(\Delta k\, l)^2 \tilde{T}}) \nonumber \\
&\propto& \exp(-\frac{m}{2 k_B T}\frac{ \left(\hbar \delta- E_R\right)^2}{ (\hbar \Delta k)^2}), \nonumber 
\end{eqnarray}
which coincides with the two-photon Doppler spectroscopy result in Eq.~\ref{eq:doppler}.

$P(q)$ provides the envelope of the sidebands, as long as assumptions (1-4) hold. In addition, in the regime of low reduced temperature $\tilde{T}$, the above results hold as long as assumption 4 is modified. In this regime, few sidebands are populated, as $k_B T \ll \hbar \omega$. In this case, the asymptotic expansion for the Bessel function only applies when $\eta^2 e^{-\tilde{\beta}/2}\gg 1$. In other words, the derived from of $P(q)$ holds as long as $\eta \gg e^{\beta/2}$. When $\tilde{T}\ll 1$, the Lamb-Dicke parameter must be exponentially larger in $1/\tilde{T}$ in order for the above results to hold.

\section{Discussion of Entropy and Temperature}
In general, entropy encodes the amount of randomness in a system. From an information theory perspective, entropy quantifies the imperfect knowledge of the precise quantum state of a system, and therefore also describes the degree of control one has over a quantum system. Therefore, motional entropy per particle, rather than absolute temperatures and phase space densities, is a more universal metric for quantifying cooling and quantum control.

In this section, we will discuss i) motional entropy for a trapped particle, ii) the relation of motional entropy to phase space density, and iii) distribution of total entropy for the specific case of a molecular tweezer array.

\subsection{Motional Entropy for a Trapped Particle}
We consider the case of a trapped particle that occupies motional eigenstates with some statistical distribution. The von Neumann entropy is given by
\begin{equation}
s=-\sum_i p_i\ln p_i,
\end{equation}
where $p_i$ is the probability that the particle is in motional state $i$. 

\subsection{Relation of Entropy to Phase Space Density}
For a classical gas at temperature $T$, the entropy per particle $s$ is given by the Sackur-Tetrode equation as
\begin{equation}
s= - \ln \left(\text{PSD}\right)+ \frac{5}{2},
\end{equation}
where $\text{PSD}= n\lambda^3$ is the phase space density, $n$ being the density and $\lambda=\sqrt{\frac{2\pi \hbar^2}{ m k_B T}}$ being the thermal de Broglie wavelength.

Quantum gases of identical bosons and fermions obey different equations of states and therefore different entropy to PSD relations $s(\text{PSD})$. These approach the classical expression at high temperature/entropies. For reference, in a homogeneous non-interacting 3D Bose gas, Bose-Einstein condensation occurs at $\text{PSD}=2.612$ with a corresponding entropy per particle of $s=1.283$. For a single component non-interacing Fermi gas in 3D, $\text{PSD} = 4/(3\sqrt{\pi} (T/T_F)^{3/2})$, where $T_F$ is the Fermi temperature. At $T/T_F=1$, $\text{PSD}=0.752$ and $s=2.95$. For our experiment, after Raman cooling, the motional entropy remains above these values at $s\approx 5$, well within the classical regime.

In terms of the cooling efficiency, the commonly used evaporative efficiency $\gamma = -d\ln (\text{PSD})/d\ln(N)$ in the classical regime can be written in terms of entropy per particle as $\gamma = ds /d\ln(N)$. This coincides with the definition $\gamma_q = ds/d\ln(N)$ defined in the main text.

\subsection{Distribution of Total Entropy in a Molecular Tweezer Array}
Here, we discuss the distribution of total entropy in the specific case of a molecular tweezer array. In addition to motional states, a molecular tweezer array contains additional degrees of freedom, in which entropy can be stored. Specifically, entropy can be stored in the spatial configuration of the molecules and the internal state distribution of the molecules. 

For the purposes of quantifying quantum control over a \textit{many-body} quantum system, one must therefore compute the total entropy per particle $s_{\text{tot}}$ stored over all these degrees of freedom. This allows a fair comparison to bulk molecular samples. In the following, we outline how to estimate $s_{\text{tot}}$ and provide a value that is achievable with our latest Raman cooling results  and previous work on initializing arbitrary 1D molecular tweezer arrays.

We assume that the different tweezer sites are uncorrelated initially. For each tweezer site, the Hilbert space contains the state $\emptyset$ corresponding to an empty tweezer without a molecule, and the states $\{\ket{i,n}\}$, corresponding to a molecule in internal state $i$ and motional state $N$ within an occupied tweezer. 

In the next sections, we provide expressions for computing the various entropy contributions.

\subsubsection{Motional Entropy}
We specifically consider a molecule trapped in a three dimensional harmonic trap with trapping frequencies $\omega_x, \omega_y$, and $\omega_z$. Assuming the occupation of the motional states follow a Boltzmann distribution with temperature $T$, the motional entropy per particle $s_m$ can be written as 
\begin{eqnarray}
	s_m&=&\sum_{\alpha=x,y,z} s_{m, \alpha}\nonumber \\
	s_{m,\alpha} &=& \bar{n}_\alpha\ln\left(\frac{\bar{n}_\alpha+1}{\bar{n}_\alpha}\right)+\ln\left(\bar{n}_\alpha+1\right),
\end{eqnarray}
where $\bar{n}_\alpha$ is the average motional quantum number along axis $\alpha$ given by
\begin{equation}
	\bar{n}_\alpha = \frac{1}{e^{\beta\hbar\omega_\alpha}-1}
\end{equation}

In our work, after 90 cycles of radial/axial Raman cooling and 30 cycles of interlaced radial cooling, $s_{m,r} =0.66(14)$, while $s_{m,z}=3.6(2)$.

\subsubsection{Configurational Entropy}
For a tweezer that is occupied with probability $p_c$, the configurational entropy for a single site is given by
\begin{equation}
\tilde{s}_c=-p_c \ln{p_c} - (1-p_c)\ln(1-p_c).
\end{equation}
For rearrangeable tweezers, recent work has shown that $p_c=0.974(1)$ can be achieved~\cite{Holland2022Tweezer}, corresponding to $\tilde{s}_c = 0.12(4)$ per site.

\subsection{Entropy of internal state distribution}
We next consider the entropy due to imperfect internal state preparation for an occupied tweezer. We denote the success rate of preparing the correct internal state as $p_{s}$. For CaF, there are 12 hyperfine states in optically cyclable $N=1$ manifold. It is experimentally challenging to measure the exact distribution over these states, but an upper bound for internal state entropy can be obtained by assuming that incorrect preparation leaves the molecule equally likely to be in the remaining 11 states.

The internal state entropy per particle $s_s$ is then given by
\begin{equation}
	s_s=-p_\text{s}\ln p_\text{s}-(1-p_s)\ln\left(\frac{1-p_s}{11}\right).
\end{equation}

In our system, using methods described in \cite{Holland2022Tweezer}, we find $p_s=0.965(17)$, which could be improved in the future. At this value, $s_s = 0.24(10)$ per particle.

\subsection{Total Entropy per Particle}
We first compute the total entropy per tweezer. Assuming no correlations between the occupations (configuration) and the motional or internal states, one can show that the total entropy per site $\tilde{s}$ is given by
\begin{equation}
\tilde{s} = \tilde{s}_c + p_s(s_m+s_i)
\end{equation}

The entropy per particle $s_{\text{tot}}$ is therefore 
\begin{equation}
	s_{\text{tot}} = \tilde{s}/p_s = \frac{\tilde{s}_c}{p_s} + s_m+s_i.
\end{equation}

One sees that at low enough fillings, configurational entropy is dominant even if all molecules are in the same internal state and in the motional ground state. For our experiment, after Raman cooling, $s_m\approx 5$ and motional entropy is the dominant contribution to the total entropy. Specifically, $s_m$ is dominated by the contribution from the axial motional degree of freedom $s_{m,z}$.

\bibliographystyle{apsrev4-2}